**RESEARCH**

**Open Access**

# OpenFOAM computational fluid dynamics (CFD) solver for magnetohydrodynamic open cycles, applied to the Sakhalin pulsed magnetohydrodynamic generator (PMHDG)

Osama A. Marzouk[1*]

*Correspondence:
Osama A. Marzouk
osama.m@uob.edu.om
[1]College of Engineering, University of Buraimi, 512 Al Buraimi, Sultanate of Oman

**Abstract**

In the current study, we present a mathematical and computational fluid dynamics (CFD) model for simulating open-cycle linear Faraday-type continuous-electrode channels of magnetohydrodynamic (MHD) power generators, operating on combustion plasma. The model extends the Favre-averaged Navier–Stokes equations to account for the electric properties of the flowing plasma gas and its reaction to the applied magnetic field. The model takes into account various effects, such as the Lorentz force, turbulence, compressibility, and energy extraction from the plasma, and it adopts an electric potential technique along with the low magnetic Reynolds number ($Re_m$) approximation. The model is numerically implemented using the multiphysics open-source computer programming environment "OpenFOAM," which combines the finite volume method (FVM) and the object-oriented programming (OOP) concept. The capabilities of the model are demonstrated by simulating the supersonic channel of the large-scale pulsed MHD generator (PMHDG) called "Sakhalin", with the aid of collected data and empirical expressions in the literature about its tested operation. Sakhalin was the world's largest PMHDG, with a demonstrated peak electric power output of 510 MW. Sakhalin operated on solid-propellant plasma (SPP), and it had a single supersonic divergent Faraday-type continuous-electrode channel with a length of 4.5 m. We check the validity of the model through comparisons with independent results for the Sakhalin PMHDG. Then, we process our three-dimensional simulation results to provide scalar characteristics of the Sakhalin channel, one-dimensional profiles along the longitudinal centerline, and three-dimensional distributions in the entire channel. For example, we show that the temperature does not change significantly along the Sakhalin PMHDG, with the outlet mass-averaged temperature being 2738.4 K, which is close to the inlet value of 2750 K. Similarly, we find that the outlet mass-averaged absolute pressure is 3.294 bar, which is near the inlet value of 3.28 bar. On the other hand, the plasma is largely decelerated from an axial speed of 2050 m/s at the inlet to 1156 m/s at the outlet (mass average). Thus, the produced pulse electric energy is primarily extracted from the kinetic energy of the plasma, rather than from its thermal energy or its pressure energy. The resolved volume-average Lorentz force density vector is [−89.12, 28.83, 0] kN/m$^3$, and the resolved volume-average electric-current density vector is [1.462, −4.517, 0] A/cm$^2$. The presented OpenFOAM solver has several applications, including preliminary design of novel geometric shapes for MHD channels, exploration of the influence of various parameters on the performance of MHD power generators (such as the inlet Mach number, the inlet pressure, and the applied magnetic-field flux density), and estimating the residual energy contained in the exit plasma for proper identification of a downstream bottoming power cycle to extract some of this available energy. Aside from the presented OpenFOAM solver, we also provide an overview of various PMHDG systems. This study can benefit different research communities, particularly those interested in OpenFOAM applications, computational fluid dynamics (CFD), magnetohydrodynamics (MHD), open-cycle MHD generators, or multiphysics mathematical modeling.

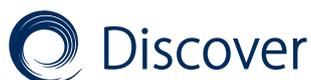

# 1 Introduction

## 1.1 Background

Sakhalin is the name of a Russian island in Northeast Asia with a conspicuously elongated form, located near the Japanese island of Hokkaido [1, 2]. Sakhalin is also the name of a powerful pulsed magnetohydrodynamic (MHD) generator that has been widely used and studied in the USSR (Union of Soviet Socialist Republics), with a demonstrated electric output of 510 MW during one of its performed tests. The Sakhalin pulsed MHD generator (PMHDG) was fueled by a solid propellant (pyrotechnic powder), and it contained a single linear continuous Faraday-type supersonic channel, with a length of 4.5 m [3–5]. The working medium for Sakhalin was hot-seeded weakly-ionized solid-propellant plasma (SPP). Continuous Faraday-type linear channels of MHG generators have two continuous electrodes placed opposite to each other. This particular linear channel design is a simpler configuration compared to other electric connection designs for linear channels [6]. Other (more-complicated) linear channel designs [7] are the segmented electrodes (segmented Faraday) design, the linear Hall design, and the diagonal design [8]. The Sakhalin was the world's most powerful PMHDG, with a theoretical peak electric output power of 600 MW (this reduces to about 400 MW with a connected matched load) [9]. It was a self-contained system, not requiring connections to external supply sources [10]. The mass of the solid propellant charge was about 6000 kg (6 tonnes). The electrodes (negative anode and positive cathode) were graphite. A possible solid propellant is a mixture of powders having 64% Mg (magnesium), 35% KNO$_3$ (potassium nitrate), and 1% technological additives (percentages are mass-based), pressed to a density of about 1700 kg/m$^3$ [11]. The propellant can also be based on metal aluminum, rather than metal magnesium [12]. A small amount of alkali metal (cesium "Cs" or potassium "K") is added to the solid propellant charge (the grain) to permit an appreciable level of ionization of the atoms of the seeded alkali metal (which changes phase to vapor at the high temperatures of the MHD channel) [13, 14]. Potassium (K) vaporizes at about 762 °C (1035 K), while cesium (Cs) vaporizes at about 682 °C (955 K) [15–17].

 Pulsed magnetohydrodynamic generators (PMHDG) can be used to produce a powerful pulse of electricity for a few seconds. PMHDG can be used in geoelectric prospecting, where the produced electric pulse allows for estimating the electric resistivity of underground formations at different depths [18–20]. PMHDG may also be used to power electromagnetic rail launchers (railguns), where projectiles can be accelerated to speeds exceeding 2.5 km/s [21, 22]. PMHDG may also find applications in space systems [23, 24] and rocket propulsion [25, 26].

Besides Sakhalin, there were other solid-propellant plasma (SPP) PMHDG in the former USSR, but Sakhalin was by large the most powerful [27–29]. A comparison of different SPP-PMHDG that existed in the former USSR is presented in Table 1 [30–32].

The table is not an exhaustive list. There were even additional SPP-PMHDG systems in the USSR, such as "Pamir" and "Pamir-06".

Aside from the pulsed mode of magnetohydrodynamic (MHD) generators, they can also operate in a continuous mode as a non-conventional topping power cycle in a dual-cycle thermal power plant [33–36]. MHD generators have a special feature of favoring elevated temperatures, which boost the electric conductivity in a rapid nonlinear fashion [37–39]. Other conventional power cycles (such as the Rankine-type steam cycle) can then be used as a bottoming cycle. In this case, the enormous unexploited heat in the exhaust hot gases leaving the topping MHD power cycle can be used to generate steam that drives steam turbines. These turbines are coupled with electromechanical generators. This dual-cycle concept is found in combined-cycle gas turbine (CCGT) power plants where a heat recovery steam generator (HRSG) is utilized to transfer heat from the higher-temperature topping gas cycle to the lower-temperature bottoming steam cycle [40–42].

The use of MHD generators, either pulsed or continuous, has an environmental advantage over fossil-fuel-fired power plants, where reducing the emissions of greenhouse gases (GHG) becomes possible, and this reduction helps in attenuating global warming [43–46]. This possibility of GHG reduction when using MHD generators arises because the fuel and oxidizer combination in the case of combustion-plasma-based MHD generators can be selected such that the carbon dioxide ($CO_2$) dominates the exploited plasma gases that are leaving the MHD generator. This carbon dioxide enrichment facilitates its capture through the oxy-fuel carbon capture (OFCC) technology [47–50]. The oxy-fuel technology is particularly attractive for MHD generators because this specialized combustion technique largely raises the temperature of the combustion flame and the combustion products. Therefore, the electric conductivity of the thermal-equilibrium "hot" plasma gas is improved. As a result, the electric output from the same volume flow rate of the plasma gas increases [51, 52]. In thermal-equilibrium "hot" plasma gas, atoms are ionized due to the temperature alone (unlike non-thermal "cold" plasma). Thus, higher temperatures increase the electric output from the same volume flow rate of the plasma gas [53, 54].

MHD generators also have the advantage of not having moving mechanical parts, which is unlike gas turbines, steam turbines, wind turbines, Stirling engines, and electromechanical generators that are commonly used in power generation [55–58]. The simplicity of the design and the lack of complex fluid–structure interaction (FSI) and solid–solid contact eliminate the need for lubrication, bearing, and sealing elements [59–61]. MHD generators are direct power extraction (DPE) systems, like solar photovoltaic (PV) modules [62] and thermoelectric generators (TEG) modules, with no intermediate energy conversion processing needed [63–66].

**Table 1** Comparison of different solid-propellant plasma (SPP) self-contained pulsed magnetohydrodynamic generators (PMHDG) in the former Union of Soviet Socialist Republics (USSR)

| Power system (ordered by power) | Pamir-0-KT | Pamir-1 | Pricaspiy | Pamir-3U | Soyuz | Ural | Khybiny | Sakhalin |
|---|---|---|---|---|---|---|---|---|
| Maximum electric power demonstrated (with an unmatched load), MW | 0.1 | 20 | 20 | 30 | 31 | 40 | 120 | 500 |
| Maximum electric power demonstrated (with a matched load), MW | – | 10 | 10 | 15 | 16 | 30 | 60 | 400 |
| Mobility | – | In containers | Mobile on a track | In containers | Mobile on a trailer | – | In containers | In containers |
| Number of MHD channels | 1 | 2 | 2 | 3 | 1 | 1 | 2 | 1 |
| Length of MHD channel(s), m | 1 | 1 | 1 | 1 | 1.8 | – | 1.17 | 4.5 |
| Operating duration for the load (duration of the pulse), s | Up to 3.6 | 3–7 | 3–7 | 2.5–10 | Up to 10 | 6 | 7–10 | Up to 7 |
| Minimum pause between two runs, h | – | 12 | 12 | – | 12 | – | 12 | 24 |
| Approximate maximum possible mass flow rate of working fluid (plasma), kg/s | 2 | 50 | 50 | 24 | 40 | 100 | 200 | 1000 |
| Overall mass of the power system, tonne | – | 8 | 12 | 18 | 25 | 16 | 34 | 50 |
| Overall dimensions of the power system, m | – | 4×1.5×2 | 4×1.5×2 | – | 7×2×2 | – | 8×7×1.5 | 13.5×3.7×2.7 |

An empty cell having a dash symbol (–) indicates an unknown value for us

One more advantage of MHD generators operating with seeded combustion plasma is their flexibility with regard to the combusted fuel. Unlike conventional internal combustion engines (ICE), either reciprocating ones used in automobiles or turbomachinery ones used in jet airplanes, the MHD generator can be viewed as an external combustion engine (ECE). This is because the combustion of the fuel occurs outside the MHD channel, which is the primary energy conversion component (extracting thermal, pressure, and/or kinetic energy from the fluid and converting the fluid energy into direct-current "DC" electricity) [67, 68]. Because the combustion products (flue gases) act mainly as a bulk high-speed medium that carries the ionized alkali metal species and the liberated charge-carrying electrons, a wide variety of fuels can be used in the plasma generator (PG) combustor unit [69–71]. For example, green hydrogen [72, 73] (hydrogen produced through water electrolysis powered by renewable energy sources) may be exploited for this type of power generation [74–77]. Gaseous fuels (such as natural gas) [78–81]. Also, liquid fuels (such as kerosene) [82–84], and solid fuels (such as coal) are candidate sources of the plasma gases [85, 86]. Despite this flexibility, gas-fired and liquid-fired MHD generators have the advantage of not forming a slag layer (a layer of molten ash material that is solidified on the walls) over the electrodes of the MHD channel [87–89].

Examining recent research work in the field of magnetohydrodynamics (MHD) shows emphasis on theoretical problems, such as MHD flow over an oscillating vertical plate in a porous medium [90], MHD natural convection [91], MHD couple stress fluid [92], MHD with a lid-driven cavity [93], and MHD for a flat plate flow [94]. Although these MHD studies contribute to our understanding of the physics of MHD principles and explore interesting academic situations, they are far from impactful implementation in the energy sector to solve environmental challenges and meet societal needs. We here address this gap by focusing on a large-scale real-world use of our MHD knowledge through novel utility-scale electricity generation in a way that admits sustainability (using green hydrogen, for example, as a fuel), compactness (limited use of space), and effectiveness (a large amount of electricity can be produced).

### 1.2 Objectives

The first goal of this work is to present a proposed computational fluid dynamics (CFD) solver for continuous-electrode (continuous-Faraday) MHD channels, which is based on the OpenFOAM object-oriented programming (OOP) environment [95]. OpenFOAM is a popular collection of open-source CFD solvers for a wide variety of problems, which include compressible flows, incompressible flows, laminar flows, turbulent flows, multiphase flows, and reacting flows [96–99]. OpenFOAM has been used in many applications in fluid dynamics [100–102], combustion [103], fluid–structure interaction (FSI) [104], and heat transfer [105]. OpenFOAM is based on the finite volume method (FVM) for discretizing the governing equations of the flow while attempting to enforce local conservation of mass, momentum, and energy at the level of a single computational cell (a single finite volume) [106, 107]. The proposed MHD channel solver in this study is a valuable tool in the computer-aided design (CAD) of MHD channels with continuous electrodes, either operating in a pulsed mode or in a sustained mode. The solver may be further extended by the interested readers to add desired design alterations or explore different operating conditions such that one or more of the performance outputs can be controlled and optimized [108, 109]. For example, the influence of the area cross-section profile of the supersonic channel (as a control variable) on the electric power output, the role of the externally applied magnetic field (applied magnetic-field flux density) on the plasma streamlines, and the use of seed alkali injection to boost the electric conductivity of the expanding plasma gas may be investigated in other studies by extending the model provided in this work [110–112]. OpenFOAM might also be viewed as an educational asset [113–115], as it can be incorporated into higher education programs related to computer engineering, fluid mechanics, numerical methods, and applied mathematics.

The second goal of this work is to present comprehensive results of three-dimensional computational fluid dynamics (CFD) modeling of the channel of the Sakhalin pulsed magnetohydrodynamic generator with solid-propellant plasma (PMHDG-SPP). Unlike various reported results in the literature, the current study shows a large number of three-dimensional and sampled one-dimensional variations of fluid fields and electric fields, as well as aggregate system-level scalar values. All these results help in understanding better the complex phenomena inside the channel. CFD modeling and post-processing of its results are very beneficial in this regard, due to the digital non-intrusive probing at any spatial location of the simulated domain. Although Sakhalin is an old system, it remains an exemplary MHD study case due to the details reported about it. Therefore, this goal is viewed as a beneficial contribution to the MHD generation field, as it can guide future attempts to construct a similar system, either pulsed for geological research or sustained for electric power generation.

The third goal of this work is to provide a brief review of former solid-propellant plasma (SPP) pulsed magnetohydrodynamic generators (PMHDG).

This study contributes to regaining interest in MHD power generation, aiming for it to be further investigated as a potential way for diversifying the energy mix locally and globally [116, 117], expanding the oxy-combustion carbon capture technology, benefiting from an exceptionally high volumetric power density [118–120], utilizing concentrated direct power extraction, facilitating the process of electrification [121, 122], and pursuing unconventional energy solutions and technology transformation [123, 124] to help in achieving net-zero emissions [125, 126].

### 1.3 Significance of the Sakhalin PMHDG within the MHD research

The Sakhalin pulsed magnetohydrodynamic generator (PMHDG) was a significant milestone in the magnetohydrodynamic (MHD) research and application because it was a full-scale (not a lab-scale or a pilot-scale) successful demonstration of the MHD concept when utilizing seeded combustion plasma. In addition, the published data about this large power unit revealed valuable sources of information about realistic operational conditions (such as what level of magnetic-field flux density is needed) and also revealed the gigantic power density that can be achieved from a relatively compact unit [127]. The adoption of the Sakhalin PMHDG in the current work as a demonstration case is an excellent example of how this system remains useful to the MHD field to date.

## 2 Research method

This section aims to describe the mathematical, computational, and geometric settings of the proposed simulation modeling for generic Faraday-type magnetohydrodynamic (MHD) channels having continuous electrodes.

By being continuous electrically-conducting electrodes (made of metal or graphite, for example), either electrode is assigned a single voltage (a single electric potential). In this study, the anode is assigned a reference voltage of zero. Thus, the "cathode voltage" here refers to the voltage difference between the higher cathode voltage and the lower anode voltage.

### 2.1 Fluid flow equations

We present here the governing partial differential equations for the plasma as a homogeneous compressible turbulent gas. These equations are extended versions of the classical Favre-averaged Navier–Stokes equations (FANS), by adding supplementary terms that mathematically account for the electromagnetic effects on the electrically-conductive plasma. The Favre-averaged Navier–Stokes equations are commonly referred to as RANS (Reynolds-averaged Navier–Stokes equations), although RANS strictly applies to incompressible (constant-density) fluids, whereas the Favre-averaged Navier–Stokes equations are an extension of RANS to compressible (variable-density) fluids, which pertain to MHD generators operating on combustion plasma [128–131]. The RANS formulation utilizes simple (unweighted) time averaging, whereas the FANS formulation utilizes density-weighted (mass-weighted) time averaging [132–134].

The scalar mass conservation equation (the continuity equation) of the plasma gas is identical to the equation governing ordinary gases (not electrically-conducting). The presence of a magnetic field and the existence of an electric conductivity of the plasma do not influence the mass conservation law of that plasma. Therefore, the mass conservation equation is

$$\frac{\partial \rho}{\partial t} + \nabla \cdot (\rho \vec{u}) = 0 \qquad (1)$$

where $\rho$ is the mass density of the plasma gas, $t$ is the time, $\vec{u}$ is the velocity vector of the plasma (this is absolute velocity, relative to a fixed laboratory frame), and $\nabla \cdot$ is the divergence operator [135, 136].

The vector linear momentum conservation equation for the plasma flow within the MHD channel is an extension of the conventional linear momentum conservation equation for a non-reacting single-phase compressible turbulent flow, where a source term is added to account for the Lorentz force per unit volume, acting on the charged particles within the plasma gas as a body force exerted on a unit volume of the plasma flowing inside the MHD channel [137, 138]. To explain the cause of this source term, it should be noted that any current-passing electrically-conducting medium that is in a translational motion while subject to a magnetic field experiences an induced force called the Lorentz force, which is perpendicular to both the translational motion and the applied magnetic field [139, 140].

Therefore, the extended plasma's linear momentum equation within the MHD channel is

$$\frac{\partial \rho \vec{u}}{\partial t} + \nabla \cdot (\rho \vec{u} \otimes \vec{u}) + \nabla p - \nabla \cdot \left( \vec{\vec{\tau}}_v + \vec{\vec{\tau}}_t \right) = \vec{J} \times \vec{B} \qquad (2)$$

where the operator $\otimes$ represents the outer product of two vectors, $p$ is the pressure (absolute pressure), $\nabla p$ is the pressure gradient vector, $\vec{\vec{\tau}}_v$ is the viscous shear stress tensor, $\vec{\vec{\tau}}_t$ is the turbulent shear stress tensor, $\vec{J}$ is the electric-current density (the current density) vector, $\vec{B}$ is the magnetic-field flux density (the magnetic field) vector, and the right-hand side term $\vec{J} \times \vec{B}$ is the added source term (the Lorentz force per unit volume of plasma) [141–144].

The viscous shear stress tensor $\vec{\vec{\tau}}_v$ is a deviatoric (traceless) tensor, modeled using the following constitutive relation [145, 146]:

$$\vec{\vec{\tau}}_v = \mu \left( \vec{\vec{\nabla}} u + \left[ \vec{\vec{\nabla}} u \right]^T - \frac{2}{3} \left( \nabla \cdot \vec{u} \right) \vec{\vec{I}} \right) \tag{3}$$

where $\mu$ is the absolute viscosity (the dynamic viscosity) of the plasma gas, $\vec{\vec{\nabla}} u$ is a tensor representing the gradient of the velocity vector, $\left[ \vec{\vec{\nabla}} u \right]^T$ is the transpose of $\vec{\vec{\nabla}} u$, and $\vec{\vec{I}}$ is the identity tensor [147, 148].

The turbulent (or Reynolds) stress tensor $\vec{\vec{\tau}}_t$ is modeled using the following gradient transport hypothesis (eddy-viscosity turbulence modeling) [149]:

$$\vec{\vec{\tau}}_t = \mu_t \left( \vec{\vec{\nabla}} u + \left[ \vec{\vec{\nabla}} u \right]^T - \frac{2}{3} \left( \nabla \cdot \vec{u} \right) \vec{\vec{I}} \right) - \frac{2}{3} \rho k \vec{\vec{I}} \tag{4}$$

where $\mu_t$ is the turbulent viscosity, $k$ is the turbulent kinetic energy per unit mass, and the spherical tensor $-\frac{2}{3} \rho k \vec{\vec{I}}$ appears due to the eddy-viscosity turbulence modeling [150, 151].

The sum of the laminar viscosity (the molecular viscosity) and the turbulent viscosity (the eddy viscosity) is the effective viscosity ($\mu_{eff}$). [152, 153]

$$\mu_{eff} = \mu + \mu_t \tag{5}$$

Therefore, the sum of the viscous stress tensor and the turbulent stress tensor can be expressed as [154, 155]

$$\vec{\vec{\tau}}_v + \vec{\vec{\tau}}_t = \mu_{eff} \left( \vec{\vec{\nabla}} u + \left[ \vec{\vec{\nabla}} u \right]^T - \frac{2}{3} \left( \nabla \cdot \vec{u} \right) \vec{\vec{I}} \right) - \frac{2}{3} \rho k \vec{\vec{I}} \tag{6}$$

It is worth mentioning that the tensor $\left( \vec{\vec{\nabla}} u + \left[ \vec{\vec{\nabla}} u \right]^T - \frac{2}{3} \left( \nabla \cdot \vec{u} \right) \vec{\vec{I}} \right)$ is the deviatoric (traceless) component of the velocity gradient tensor $\vec{\vec{\nabla}} u$, which has a trace of $\frac{2}{3} \left( \nabla \cdot \vec{u} \right)$, and this trace quantity vanishes for incompressible fluids, but it is not zero here in the presented MHD plasma low model [156–159].

A transport scalar partial differential equation is solved for obtaining the spatial distribution and temporal evolution of the turbulent kinetic energy per unit mass ($k$), and another scalar partial differential equation is solved for obtaining the dissipation rate ($\epsilon$ or epsilon) [160, 161].

The scalar equation of energy conservation is written here using the total (stagnation) specific internal energy, $\tilde{e}$, as the dependent variable, instead of the temperature $T$. This form is suitable for high-speed gases, as encountered in open-cycle MHD channels [162–165]. As with the case of the vector momentum equation, the scalar energy equation has a source term that is not present in the ordinary form (for non-electrically-conductive gases). This source is the power (per unit volume of plasma) extracted from the plasma and converted into electric power (per unit volume of plasma) to an external electric load connected to the electrodes of the MHD channel.

Thus, the extended energy equation for the plasma gas is

$$\frac{\partial \rho \tilde{e}}{\partial t} + \nabla \cdot (\rho \vec{u} \tilde{e}) + \nabla \cdot (\kappa_{eff} \nabla T) + \nabla \cdot \left(\vec{u} \cdot \vec{\vec{\tau}}\right) = \vec{J} \cdot \vec{E}_L \quad (7)$$

where $\kappa_{eff}$ is the effective thermal conductivity (accounting for both laminar/molecular effect and turbulent/eddy effect in heat conduction), $-\kappa_{eff} \nabla T$ is the effective diffusion heat flux vector (expressed in terms of the temperature gradient $\nabla T$ according to Fourier's law), $\nabla \cdot \left(\vec{u} \cdot \vec{\vec{\tau}}\right)$ is the viscous-stresses power (rate of work by the plasma) per unit plasma volume due to viscous stresses, and $\vec{J} \cdot \vec{E}_L$ is the source term due to the electric properties of the plasma [166, 167]. The vector $\vec{E}_L$ is the electrostatic field (caused by the voltage difference between the MHD electrodes due to the externally applied load "if applicable" or due to the external electric circuit being open "the electrodes are not electrically connected electrically"). The vector field $\vec{E}_L$ is an applied electric field on the plasma due to the voltage difference at the electrodes surrounding it, and these electrodes (grounded anode and positive cathode) are used to power an external electric load, thus we also refer to this electric field as the electrostatic field. The vector field $\vec{E}_L$ is not directly dependent on the motion of the plasma, which explains the name "electrostatic". The vector field $\vec{E}_L$ can even be externally applied and ma manually controlled, by connecting the electrodes of the MHD channel to a DC power source (instead of an electric load). On the other hand; another electric field is induced due to the translational motion of the electric-conductive plasma gas with a velocity vector $\vec{u}$ within an applied magnetic magnetic-field flux density $\vec{B}$; this induced electric field is $\vec{u} \times \vec{B}$, and it can be referred to as an "electrodynamic" field. The load electric field $\vec{E}_L$ is irrotational (as per Coulomb's law) [168, 169]. This means

$$\nabla \times \vec{E}_L = 0 \quad (8)$$

The source term $\vec{J} \cdot \vec{E}_L$ actually has a negative value (the electrostatic field vector is directed against the electric-current density vector), making this source term act actually as an energy sink, which is the valid interpretation of it as an expression of the process of extracting energy from the plasma flow and supplying it as electric energy to an external electric load. In the idealized (most simplified) case of one-dimensional (unidirectional) constant plasma velocity through a constant-cross section MHD channel, one-dimensional (unidirectional) constant applied magnetic field and one-dimensional (unidirectional) constant electrostatic field being perpendicular to the parallel electrodes (from the grounded anode to the positive cathode), and with the three fields are mutually orthogonal to each other; this energy source term reduces to $-JE_L$. Thus, it is actually an energy sink term with a magnitude of $|JE_L|$, where $J$ and $E_L$ reduce to scalar values (but the current density $J$ is from the anode to the cathode, while the electrostatic field $E_L$ is from the cathode to the anode) [170].

The total specific internal energy $\tilde{e}$ is the sum of the static specific internal energy $e$ and the specific kinetic energy $0.5 \vec{u} \cdot \vec{u}$. Mathematically, this is expressed as [171]

$$\tilde{e} = e + \frac{1}{2}\vec{u} \cdot \vec{u} \tag{9}$$

The static specific internal energy $e$ is related to the static specific enthalpy $h$ as [172, 173]

$$h = e + p/\rho \tag{10}$$

Likewise, the total specific internal energy $\tilde{e}$ is related to the total specific enthalpy $\tilde{h}$ as [174, 175]

$$\tilde{h} = \tilde{e} + p/\rho \tag{11}$$

This means

$$\tilde{h} = h + \frac{1}{2}\vec{u} \cdot \vec{u} \tag{12}$$

The ideal gas law (ideal gas equation of state) is also needed in the MHD model, as an algebraic relation between the absolute pressure $p$, absolute temperature $T$, and density $\rho$. The ideal gas law is not affected by the electric characteristics of the weakly-ionized seeded combustion plasma, which is still treated as a gas thermodynamically. The ideal gas law has the form [176–178]

$$p = \rho RT \tag{13}$$

where $R$ is the specific gas constant for the plasma gas, which is related to the universal (molar) gas constant $\overline{R}$ and the plasma molecular weight $W$ as [179]

$$R = \overline{R}/W \tag{14}$$

The value of $\overline{R}$ is 8.314 J/(mol.K) or 8,314 J/(kmol.K) [180]. If the plasma is treated as a mixture of two or more gaseous species, then the molecular weight $W$ depends on the local chemical composition of the plasma (at a given computational cell) as [181, 182]

$$W = \sum_i^N X_i W_i \tag{15}$$

where $X_i$ is the mole fraction of the i[th] gaseous species in the plasma gas mixture (such as carbon dioxide, $CO_2$; or water vapor, $H_2O$), $W_i$ is the molecular weight of that ith gaseous species, and $N$ is the number of gaseous species in the plasma gas mixture.

### 2.2 Electric fields equations

In the previous subsection, it was shown that three electromagnetic vector fields are needed in order to integrate the system of partial differential equations governing the density, velocity, and the total specific internal energy of the plasma flow within the MHD channel. These three electromagnetic vector fields appear in the two source terms that arise in the extended vector momentum equation and the extended scalar energy equation. These three needed electromagnetic vector fields are (1) the electric-current density vector (or simply the current density) $\vec{J}$, (2) the magnetic-field flux density vector (or simply the magnetic field) $\vec{B}$, (3) and the electrostatic vector field $\vec{E}_L$.

The magnetic-field flux density $\vec{B}$ is a specified applied field. Thus, in the proposed MHD model here, only the current density field $\vec{J}$ and the electrostatic field $\vec{E_L}$ need to be estimated.

The approach proposed here for handling the electric aspect of the problem utilizes a scalar electric potential $\phi$ field as a working variable that is directly solved for, and then its solution is used to indirectly compute the fields $\vec{J}$ and $\vec{E_L}$.

The electric potential scalar field $\phi$ is related to the electrostatic vector field $\vec{E_L}$ as follows:

$$\vec{E_L} = -\nabla \phi \tag{16}$$

Thus, the gradient of the electric potential is the electrostatic field, but in the opposite direction. The above mathematical relation is made possible given that the electrostatic field $\vec{E_L}$ is irrotational [183–186].

The electric potential $\phi$ is governed by a Poisson-type parabolic partial differential equation as [187, 188]

$$\nabla \cdot \left( \vec{\sigma_{eff}} \cdot \nabla \phi \right) = \nabla \cdot \left( \vec{\sigma_{eff}} \cdot \left( \vec{u} \times \vec{B} \right) \right) \tag{17}$$

where $\vec{\sigma_{eff}}$ is an effective electric conductivity tensor for the plasma gas. For the case of unidirectional magnetic-field flux density in the $Z$ direction only (thus, $\vec{B} \equiv [0, 0, B]$), with $B$ being the magnitude of that magnetic-field flux density (which can be a function of space rather than being a uniform constant), the effective electric conductivity tensor has the following form:

$$\vec{\sigma_{eff}} = \sigma \begin{bmatrix} 1 & -\frac{\beta}{\beta^2+1} & 0 \\ \frac{\beta}{\beta^2+1} & 1 & 0 \\ 0 & 0 & 1 \end{bmatrix} \tag{18}$$

where $\sigma$ is the scalar electric conductivity of the plasma, and $\beta$ is the dimensionless Hall parameter (electron's Hall parameter), which is the product of the electron mobility $\mu_e$ and the magnitude of the applied magnetic field $B$ [189–191]. Therefore,

$$\beta = \mu_e B \tag{19}$$

The effective electric conductivity tensor is used to express the generalized form (the vector form) of Ohm's law for the electrically-conductive plasma, as [192, 193]

$$\vec{J} = \vec{\sigma_{eff}} \cdot \left( \vec{E_L} + \vec{u} \times \vec{B} \right) \tag{20}$$

It should be noted that the vector $\vec{E_L} + \vec{u} \times \vec{B}$ is the net electric field acting internally on the plasma (as an electrically-conductive medium having its own internal electric resistivity), and it accounts for both the electrostatic field $\vec{E_L}$ and the electrodynamic field $\vec{u} \times \vec{B}$. As an illustrative analogy with a simple battery-lamp circuit, the field $\vec{u} \times \vec{B}$ is likened to the battery voltage, the field $\vec{E_L}$ is likened to the voltage drop across the lamp, and the field $\vec{E}_L + \vec{u} \times \vec{B}$ is likened to the voltage drop across the battery itself due to its internal electric resistance. Furthermore, the absolute value of the energy equation source term $\left|\vec{J} \cdot \vec{E}_L\right|$ can be likened to the power dissipated in the lamp. In the plasma's energy equation, this term has a negative value because, for the plasma, it is power extracted (removed).

It should also be noted that the proposed solution method for the electric fields assumes that the magnetic-field flux density $B$ is fixed in time (but it does not need to be fixed in space), and thus the magnetic-field flux density $B$ is not affected by the plasma flow. This corresponds to the low magnetic Reynolds number approximation [194, 195]. The magnetic Reynolds number ($Re_m$) is a dimensionless quantity that helps in quantifying the interaction between the moving plasma and the magnetic field [196, 197]. Under large magnetic Reynolds numbers ($Re_m > > 1$), the magnetic field lines can be likened to elastic bands that are frozen into the electrically-conductive plasma [198]. The velocity of the plasma establishes a secondary induced magnetic-field flux density $\vec{b}$, which is added to the applied magnetic-field flux density $\vec{B}$, making the total magnetic-field flux density actually dependent on the plasma flow. This additional flow-dependent magnetic-field flux density then affects the plasma velocity through the source term in the plasma momentum equation, presented in Eq. (2). Thus, at high magnetic Reynolds numbers, a two-way coupling exists between the magnetic-field flux density seen by the traveling plasma and the velocity field of that plasma. On the other hand, under small magnetic Reynolds numbers ($Re_m < < 1$), the movement of the plasma has a negligible effect on the magnetic-field flux density acting on it. Therefore, it is reasonable in that case to treat the magnetic-field flux density acting upon the plasma as the fixed externally-applied $\vec{B}$ field only. This means that the low magnetic Reynolds number approximation treats the coupling between the magnetic-field flux density and the plasma motion as having a one-way direction, with the externally-applied magnetic-field flux density affecting the motion of the plasma while the induced current density within the moving plasma does not distort the magnetic field through producing an additional induced one [199]. The low magnetic Reynolds number (low-$Re_m$) approximation is also called an "inductionless" approximation [200]. On the other hand, when the induced magnetic-field flux density is accounted for, this corresponds to a "self-excitation" or "self-excited" system [201].

Mathematically, the magnetic Reynolds number ($Re_m$) is defined as

$$Re_m = \mu_m^* \sigma^* u^* L^* \tag{21}$$

where $\mu_m^*$ is a reference magnetic permeability. The magnetic permeability of vacuum (the permeability of free space or the magnetic constant) $\mu_0$ is a universal constant with the exact value of $4\pi \times 10^{-7}$ H/m [202–204]. In the above equation defining $Re_m$, $\sigma^*$ is a reference electric conductivity of the electrically-conductive medium, $u^*$ is a reference scalar speed for the electrically-conductive medium, and $L^*$ is a reference length relevant to the geometric scales of the problem.

The definition of the magnetic Reynolds number ($Re_m$) in Eq. (21) can be reformulated as

$$Re_m = \frac{u^* L^*}{\mu_m^* \sigma^*} \tag{22}$$

where the product $\mu_m^* \sigma^*$ has the dimension of (area/time), which is the same as the momentum diffusivity (the kinematic viscosity) for fluids. Therefore, the quantity $\mu_m^* \sigma^*$ may be called a reference "magnetic diffusivity", and assigned the symbol $\nu_m^*$ that resembles the symbol $\nu^*$ used for designating a reference momentum diffusivity (a reference kinematic viscosity) in fluid mechanics [205].

Equation (22) thus can be written as

$$Re_m = \frac{u^* L^*}{\nu_m^*} \tag{23}$$

The above form is very similar to the one used for defining the conventional Reynolds number (Re) in fluid mechanics applications, which is [206, 207]

$$Re = \frac{u^* L^*}{\nu^*} \tag{24}$$

This conventional Reynolds number (Re) in fluid mechanics is used as a nondimensional criterion to infer whether the fluid flow is expected to be in a laminar (regular and smooth) regime or a turbulent (irregular with a lot of mixing) regime. However, unlike the magnetic Reynolds number ($Re_m$); the threshold of regime transitioning for the conventional Reynolds number (Re) tends to be much larger than unity, and this threshold is dependent on the type of flow (for example, internal flows inside a cylindrical pipe have a lower threshold than external flows over the same cylindrical pipe).

The low magnetic Reynolds number (low-$Re_m$) approximation is relevant in MHD generators, and it is adopted in the proposed MHD model here [208, 209]. To demonstrate this relevance, we consider a representative situation of a weakly-ionized plasma gas for MHD channels, formed by seeding hot air at 3000 °C (3,273 K) with a small amount of potassium vapor (2% fraction). Under a reasonable reference length of 1 m and a reasonable reference speed of 1000 m/s, the magnetic Reynolds number ($Re_m$) is near 0.13 only, while the conventional Reynolds number (Re) is about $3.5 \times 10^6$ [210]. For the Sakhalin pulsed MHD generator, the induced magnetic-field flux density $\vec{b}$ is small compared to the external magnetic-field flux density $\vec{B}$, with the peak of $\vec{b}$ (about 0.19 T) being less than 10% of the peak of $\vec{B}$ (about 2.1 T). This induced magnetic-field flux density causes the overall magnetic-field flux density (the sum of $\vec{B}$ and $\vec{b}$) to be less than the externally applied $\vec{B}$ at the first region of the channel (nearly in its first half), but more than the externally applied $\vec{B}$ in the second region of the channel (nearly in its second half), and the alteration due to $\vec{b}$ are smallest near the middle of the longitudinal dimension of MHD channel (around $X$ = 2.25 m). The influence of the induced $\vec{b}$ on the estimated electric power generation is only 2.5% [211]. So, our low magnetic Reynolds number (low-$Re_m$) approximation is considered suitable, despite that the magnetic Reynolds number ($Re_m$) for the Sakhalin PMHD generator is not very small compared to unity, but it is still less than one (about 0.5).

For a solid conductor, the generalized form (vector form) of Ohm's law, as presented in Eq. (20), reduces to the classical scalar form of Ohm's law, which is

$$I = V/R \qquad (25)$$

where $I$ is the electric current, $V$ is the voltage difference (voltage drop across the electric load), and $R$ is the electric resistance of the electric load.

Using Eq. (18) in Eq. (20), and expanding the resulting expression from a vector equation into three scalar equations gives the following three Cartesian components of the electric density vector, $\vec{J} \equiv [J_x, J_y, J_z]$:

$$J_x = \frac{\sigma}{\beta^2 + 1} \left[ (E_{L,x} + vB) - \beta (E_{L,y} - uB) \right] \qquad (26)$$

$$J_y = \frac{\sigma}{\beta^2 + 1} \left[ (E_{L,y} - uB) + \beta (E_{L,x} + vB) \right] \qquad (27)$$

$$J_z = \sigma E_{L,z} \qquad (28)$$

where $u, v,$ and $w$ are the three Cartesian components of the plasma velocity vector $\vec{u}$, and $E_{L,x}, E_{L,y},$ and $E_{L,z}$ are the three Cartesian components of the electrostatic vector $\vec{E_L}$. Thus, $\vec{u} \equiv [u, v, w]$. and $\vec{E_L} \equiv [E_{L,x}, E_{L,y}, E_{L,z}]$. It can be noticed that while there is a coupling between $J_x$ and $J_y$, they are decoupled from $J_z$. This decoupling is a result of assuming the magnetic-field flux density $\vec{B}$ to be unidirectional, thus $\vec{B} \equiv [0, 0, B]$ in the proposed MHD model.

The Poisson equation, Eq. (17), for the electric potential $\phi$ has the electric potential in the left-hand side only, which is treated with implicit discretization, while the right-hand side has the flow velocity, which is discretized explicitly.

When the electric potential $\phi$ is obtained by integrating the Poisson equation, Eq. (17), numerically, a Dirichlet boundary condition (a first-type boundary condition) is applied at the anode (the top MHD boundary), by enforcing the value of $\phi$ to be zero, which is an arbitrary value that represents electric grounding [212–214]. Another Dirichlet boundary condition is applied at the positive cathode electrode (the bottom MHD boundary), by enforcing the value of $\phi$ there to be equal to the aimed value for the MHD operation (the target voltage drop needed for powering the external electric load, or the target open-circuit voltage if no load is connected). Therefore, the two boundary conditions for the electric potential $\phi$ at the two electrodes are

$$\phi_{anode} = 0 \qquad (29)$$

$$\phi_{cathode} = \Delta V_L \qquad (30)$$

where $\Delta V_L$ is the target voltage difference across the electrodes and thus across the externally connected load (or the open-circuit voltage).

For the two other boundaries of the MHD channel (the non-electrically-conductive walls), a Neumann boundary condition (a second-type boundary condition) is applied, which has the form [215–217]

$$\frac{\partial \phi}{\partial n} (or \nabla \phi \cdot \hat{n}) = \left( \vec{u} \times \vec{B} \right)_{boundary} \cdot \hat{n} \qquad (31)$$

where $\hat{n}$ is a unit normal vector pointing perpendicularly away from the MHD boundary wall (as if leaving the channel laterally).

After obtaining the electric potential field $\phi$, the electrostatic field $\vec{E}_L$ is computed as the negative gradient of $\phi$ using Eq. (32), which is repeated below for a convenient flow of the analysis.

$$\vec{E}_L = -\nabla \phi \tag{32}$$

Finally, the generalized Ohm's law is used to compute the electric-current density $\vec{J}$ according to Eq. (33), which is repeated below for a convenient flow of the analysis.

$$\vec{J} = \vec{\sigma}_{eff} \cdot \left( \vec{E}_L + \vec{u} \times \vec{B} \right) \tag{33}$$

### 2.3 Geometry and mesh of the Sakhalin MHD channel

The geometry and operational settings for the modeled Sakhalin MHD channel are illustrated in Fig. 1. The bulk flow direction of the supersonic plasma is along the $X$-axis. The unidirectional magnetic field depends only on the axial distance ($X$) from the channel inlet, and the range of $X$ is from 0 to 4.5 m. The anode is the upper surface of the channel (having positive linearly-increasing $Y$ coordinates), while the cathode is the lower surface of the channel (having negative linearly-decreasing $Y$ coordinates). The width of the channel (along the $Z$ direction) is constant, with a value of 1 m. The height (along the $Y$ direction) of the channel increases from 0.9 m at the inlet to 1.6 m at the outlet.

Projected views (front view and top view) of the MHD channel are provided in Fig. 2. The front view is a trapezium (a trapezoid), while the top view is a rectangle.

The spatial discretization of the Sakhalin MHD channel is illustrated in Fig. 3. Due to its geometric simplicity, the channel is discretized spatially as a single block, with a total of 376,380 hex cells (hexahedral computational cells). We used 180 cells along the axial $X$ dimension (the length), 51 cells along the vertical $Y$ dimension (the height), and 41 cells along the lateral $Z$ dimension (the width). Hexahedral (structured) meshes tend to be favored over tetrahedral (unstructured) meshes due to the higher accuracy and the ability to have a larger aspect ratio [218–220]. Similarly, in two-dimensional problems, quadrilateral cells are computationally preferred over triangular cells [221, 222]. Therefore, when possible, a tetrahedral mesh may be limited to complex geometries that cannot be represented well through hexahedral mesh elements. In our CFD simulation of the Sakhalin MHD channel, tetrahedral cells were not necessary.

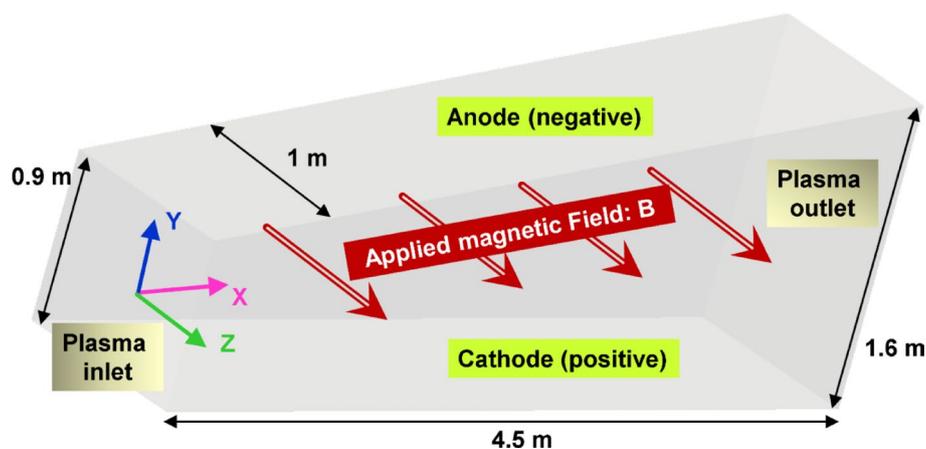

**Fig. 1** Three-dimensional illustration of the modeled Sakhalin MHD channel

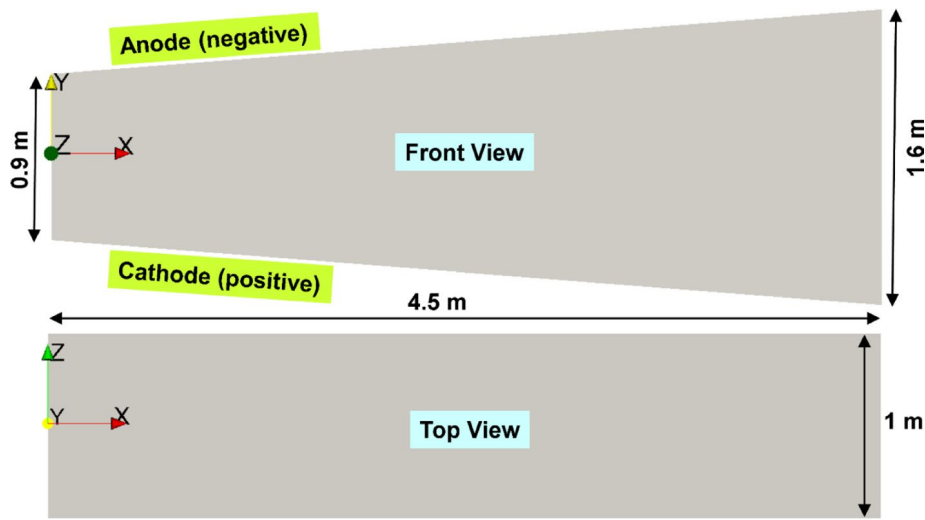

**Fig. 2** Projection views of the modeled Sakhalin MHD channel

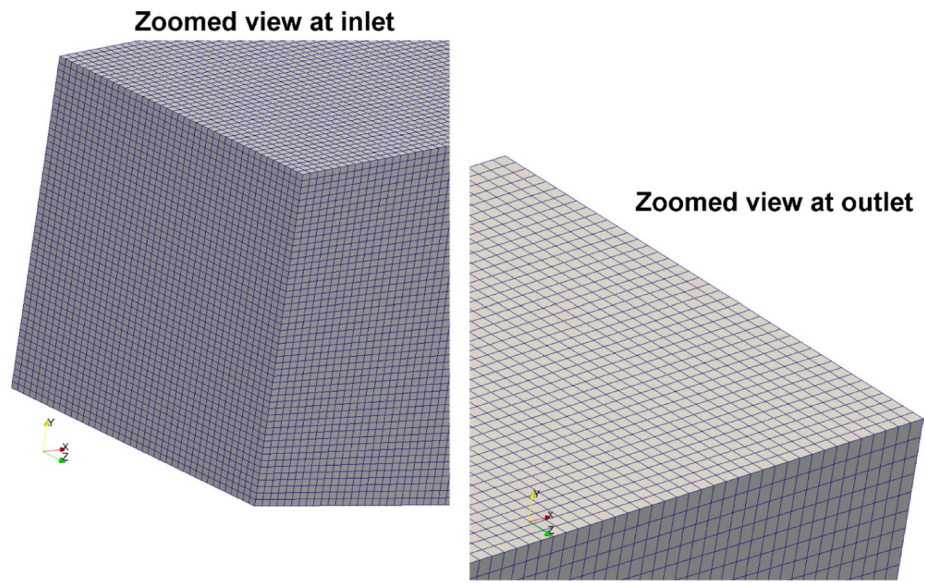

**Fig. 3** Zoomed views of the discretization mesh of the Sakhalin MHD channel

Our selected spatial discretization makes each cell nearly a cube (cells have a comparable size of their edges, near 0.025 m), and this rough regularity in the mesh is preferred in CFD simulations compared to highly skewed or stretched cells [223, 224]. Furthermore, our selection of an odd number of cells for the height (51 cells) and the width (41 cells) ensures that the longitudinal centerline of the channel coincides with a sequence of cell centers, which are also equally spaced. This is helpful because we sample the simulation results along that longitudinal centerline, which starts from the $[X, Y, Z]$ coordinates [0,0,0] and ends at the point [4.5,0,0]. Thus, exact cell-center values are readily available for sampling, without approximating interpolation.

### 2.4 Empirical expressions for Sakhalin

The previous mathematical description of the governing equations and computational modeling for the MHD plasma is not restricted to the Sakhalin pulsed magnetohydrodynamic generator (PMHDG). However, a number of thermofluidic and electromagnetic properties appear in these governing equations, which should be modeled in the computational fluid dynamics (CFD) simulations using customized submodels [225–227]. Such properties include the specific heat capacity at constant pressure $C_p$, the magnitude of the unidirectional magnetic-field flux density $B$ the Hall parameter $\beta$, and the scalar electric conductivity $\sigma$.

The Sakhalin PMHDG has empirical expressions published in the literature that describe the aforementioned four quantities [228]. Such expressions are listed in Table 2. It should be noted that they correspond to the Pamir-3U SPP-PMHDG. No expressions for Sakhalin itself were available. However, because such data were published. Due to the similarity in the operation between Pamir-3U and Sakhalin (but Sakhalin has a much larger power capacity), these expressions were assumed to be valid also for Sakhalin. In addition, the same table lists other operational conditions used in our simulation based on published data in the literature.

For the magnitude of the magnetic-field flux density $B$, the expression listed in the table is our seventh-degree polynomial fitting, which we obtained using the (curve_fit) function of the (optimize) package for optimization and root finding, provided with the (SciPy) open-source Python library [229–231]. The function (curve_fit) utilizes a nonlinear least squares method. Our proposed fitting function is visualized in Fig. 4.

### 2.5 Computational Settings and Demands

As with other computational fluid dynamics (CFD) simulations, a number of choices need to be made with regard to discretization schemes and/or numerical parameters. In this subsection, we cover this aspect of the OpenFOAM model presented here for magnetohydrodynamics (MHD) channels by listing some configuration settings adopted in the simulation in Table 3.

The simulated time of 0.04 s was selected because the solution reached a quasi-steady state, with negligible transience. Although the governing equations are unsteady, the simulation solution reached a stable stage.

The density ($\rho$) and the conservative form of the velocity vector ($\rho \vec{u}$) are resolved using a diagonal solver (thus, solved explicitly) [245, 246]. The total specific internal energy ($\tilde{e}$) is resolved using a Gauss Seidel Smoother [247]. The turbulent kinetic energy per unit mass ($k$) and the turbulence dissipation rate ($\epsilon$ or epsilon) are resolved using a preconditioned bi-conjugate gradient (PBiCG) solver [248], with a Diagonal-based Incomplete Lower–Upper decomposition preconditioner (DILU) [249, 250].

We used a computing machine having two processors. The type of these processors is quad-core Intel Xeon L5335 2.00 GHz. Thus, the computing machine has a total of eight computing cores. One simulation can take one full day (24 h). The simulation environment was the Linux operating system (the openSUSE distribution).

### 2.6 OpenFOAM Code Lines

We provide below code lines from the proposed OpenFOAM magnetohydrodynamic (MHD) solver, based on our mathematical description provided earlier.

**Table 2** Empirical expressions and operating conditions for Sakhalin plasma

| Quantity | Unit | Expression or value | Remarks |
|---|---|---|---|
| Specific heat capacity at constant pressure, $C_p$ | J/(kg.K) | $C_p T = 649.863 + 0.620932\,T$ | $C_p$ increases linearly with the temperature<br>$T$ is in kelvins (absolute temperature) |
| Magnetic-field flux density magnitude, $B$ | T (tesla) | $B(X) = \sum_{i=0}^{7} a_i X^i$<br>$a_0$=1.59103012,<br>$a_1$=1.62627541,<br>$a_2$=−2.41601024,<br>$a_3$=1.81448629,<br>$a_4$=−0.777421291, $a_5$=0.188450611,<br>$a_6$=−0.0230846554,<br>$a_7$=0.00110429100 | Our seventh-degree polynomial fit of a published curve<br>$X$ is the axial distance (in meters) from the inlet of the MHD channel<br>$X$ starts from 0 at the MHD channel inlet<br>$X$ reaches 4.5 m at the MHD channel outlet |
| Hall parameter, $\beta$ | – | $\beta = B\dfrac{b_0 + b_1\sqrt{T}}{\widetilde{p}}$<br>$b_0$=0.393664,<br>$b_1$=0.00302522 | $B$ is in teslas<br>$T$ is in kelvins (absolute temperature)<br>$\widetilde{p}$ is the absolute pressure expressed in standard atmospheres (atm = 101,325 Pa) |
| Scalar electric conductivity, $\sigma$ | S/m (siemens per meter) | $\sigma(\widetilde{p},T) = c_0 \exp\left(\dfrac{c_1 + c_2 T}{\widetilde{p}} + c_3 + c_4 T + (c_5 + c_6 T)\widetilde{p}\right)$<br>$c_0$ = 1.15184,<br>$c_1$ = −3.29702,<br>$c_2$ = 0.00140816,<br>$c_3$ = −2.02024,<br>$c_4$ = 0.00212241,<br>$c_5$ = −0.156856,<br>$c_6$=0.0000319339 | $T$ is in kelvins (absolute temperature)<br>$\widetilde{p}$ is the absolute pressure expressed in standard atmospheres (atm = 101,325 Pa) |
| Cathode voltage, $\phi_{\text{cathode}}$ | V (volt) | 2,550 | The anode voltage is $\phi_{\text{anode}} = 0$<br>This value corresponds to the experimental "run number 1" of the MHD channel |
| Plasma molecular weight, $W$ | kg/kmol | 22.905 | $W$ is assumed to be uniform |
| Plasma specific gas constant, $R$ | J/(kg.K) | 362.978 | $R = 8{,}314/W$ |
| Inlet temperature, $T_{in}$ | K (kelvin) | 2,750 | 2750 K = 2,77 °C |
| Inlet pressure, $p_{in}$ | bar | 3.28 | 3.28 bar = 328,000 Pa = 3.2371 atm |
| Inlet axial velocity, $u_{in}$ | m/s | 2050 | 2050 m/s = 7380 km/h |
| Inlet specific heat capacity at constant pressure, $C_{p,in}$ | J/(kg.K) | 2357.426 | $C_{p,in} = C_p(T = T_{in})$ |
| Inlet specific heat capacity at constant volume, $C_{v,in}$ | J/(kg.K) | 1994.448 | $C_{v,in} = C_{p,in} - R$ (ideal gas [232]) |
| Inlet specific heat ratio, $\gamma_{in}$ | – | 1.1820 | $\gamma_{in} = C_{p,in}/C_{v,in}$ [233] |
| Inlet speed of sound, $a_{in}$ | m/s | 999.5 | $a_{in} = \sqrt{\gamma_{in} R T^*}$ [a] (ideal gas [234]) |
| Inlet Mach number, $M_{in}$ | – | 2.05 | $M_{in} = u_{in}/a_{in}$ [235] |
| Inlet Plasma density, $\rho_{in}$ | kg/m³ | 0.32859 | $\rho_{in} = p_{in}/(R T_{in})$ |
| Inlet cross-section area, $A_{in}$ | m² | 0.9 | Width = 1 m, height 0.9 m |
| Mass flow rate, $\dot{m}$ | kg/s | 606.2 | $\dot{m} = u_{in} \rho_{in} A_{in}$ [236] |
| Outlet cross-sectional area, $A_{out}$ | m² | 1.6 | Width = 1 m, height 1.6 m |

[a] This speed of sound corresponds to a temperature of $T^*$ = 2,328 K. If the centerline plasma temperature of $T_{in}$ = 2,750 K is used, the formula gives 1086.2 m/s. The temperature at the inlet can be viewed as a combined effect of the hotter core plasma and the colder walls and near-wall region, or as the temperature of the plasma at the centerline, which has the highest temperature in the inlet boundary

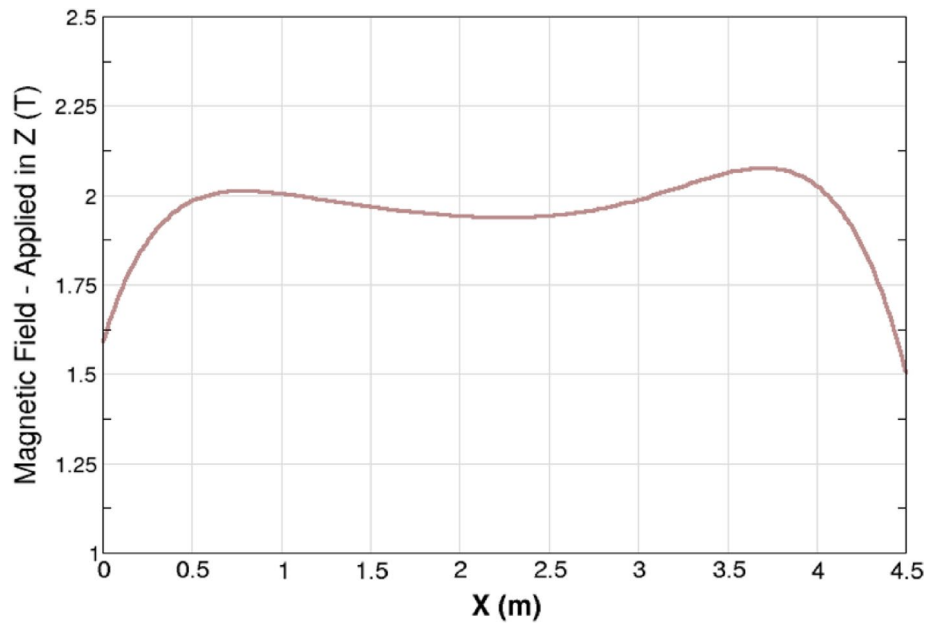

**Fig. 4** Nonlinear polynomial fitting (seventh-degree polynomial) for the magnitude of the magnetic-field flux density (***B***) of the Sakhalin MHD channel, as a function of the axial distance (***X***). The unidirectional magnetic field is in the ***Z*** direction, and is expressed in teslas (***T***)

**Table 3** Some simulation conditions for the CFD simulation of the Sakhalin mhd channel

| Condition | Adopted setting |
| --- | --- |
| Turbulence model | Standard k-epsilon [237, 238] |
| Flow solver | Density-based, nearly explicit |
| Upwind scheme | Kurganov central [239–242] |
| Prandtl number (Pr) | 0.72 [243, 244] |
| Time-step | $5 \times 10^{-6}$ s (5 μs) |
| Completed time steps | 8000 |
| Simulated time | 0.04 s |
| Convergence tolerance | $10^{-10}$ |

For example, the first part of these code lines shows how the Poisson equation, Eq. (17), for the electric potential $\phi$ is solved using the high-level OpenFOAM programming language, through an iterative loop with (nCorr) iterations. In the code, the electric potential $\phi$ is assigned the variable name (phi_e).

The remaining parts of the code lines show how the scalar mass conservation equation is solved, followed by the vector momentum equation, and finally the scalar energy equation [251].

```cpp
for (int i = 0; i < nCorr; i++)
    {
      fvScalarMatrix phiEqn
      (
        fvc::div( sigmaEff & (U ^ B) ) - fvm::laplacian(sigmaEff,phi_e)
      );
      phiEqn.relax();
      phiEqn.solve();
      E = -fvc::grad(phi_e);
      J = sigmaEff & ( E + (U^B) );
      lorentzForceDensity = J ^ B;
      powerDensity = E & J;
    }
solve(fvm::ddt(rho) + fvc::div(phi));
solve
    (
      fvm::ddt(rho, U) - fvc::ddt(rho, U)
      - fvm::laplacian(muEff, U)
      - fvc::div(tauEffExplicit)
    );
surfaceScalarField TauEffDotUDotSf
    (
      (
        fvc::interpolate(muEff)*mesh.magSf()*fvc::snGrad(U)
       + (mesh.Sf() & fvc::interpolate(tauEffExplicit))
      )
      & (frac_ap*U_ifAllFluxesOut + frac_am*U_ifAllFluxesIn)
    );
solve
    (
      fvm::ddt(rhoE)
      + fvc::div(phiE_pU)
      - fvc::div(TauEffDotUDotSf)
      - powerDensity
    );
e = rhoE/rho - 0.5*magSqr(U);
e.correctBoundaryConditions();
volScalarField kEff("kEff", thermo.Cp()*turbulence->alphaEff());
solve
    (
      fvm::ddt(rho, e) - fvc::ddt(rho, e)
      - fvc::laplacian(kEff, T)
    );
rhoE = rho*(e + 0.5*magSqr(U));
```

### 2.7 Limitations

Despite the broad range of physical, chemical, electromagnetic, and gas dynamic phenomena covered in the presented magnetohydrodynamic (MHD) solver for linear channels of MHD generators, a number of aspects are not captured. Although this means limitations in the solver, this does not contradict the usefulness and importance of the solver. Such limitations may be addressed by extending the solver according to the customized setting and purpose of its implementation. In a setting where the ignored details are not strongly present, the solver becomes adequate.

For example, if the fuel combustion results in a two-phase (solid particle carried by a bulk gaseous medium) with a large fraction of the solid phase that cannot be neglected, then an extension may be necessary to take into account the influence of these particles. However, fuels not featuring such behavior do not require such an extension of the solver.

As another example, in the introduced MHD solver, we acknowledge that electrode erosion is overlooked. This means that the electrodes are assumed to remain intact during the MHD channel operation. This assumption is not a concern for performance analysis, since erosion becomes noticeable only after a large period of operation. Thus, ignoring its effect is not a shortcoming, but it is still a possible area of improvement.

## 3 Validation and verification

After presenting the mathematical/computational formulation of the OpenFOAM solver developed for handling continuous Faraday channels in MHD generators, this section is provided to compare some results we obtained using the solver with values previously published in the literature, and this is a validation step for our CFD solution of the Sakhalin pulsed magnetohydrodynamics generator (PMHG), through external benchmarking. In the verification step, we perform internal benchmarking by comparing the CFD solution at the nominal spatial and temporal resolutions with another CFD solution using the same solver but at finer spatial and temporal resolutions; this verifies the ability of the CFD model to converge at the nominal resolutions [252, 253].

Before presenting the validation step; we point out that due to the uncertainty in reported data of the Sakhalin PMHDG where a range for a variable is reported, rather than a single value for it (such as the mass flow rate), the discrepancy of some reported operational characteristics (such as the MHD channel inlet velocity), the time-unsteadiness nature of the actual problem, our simplification introduced in the inlet conditions, and the empirical fitting functions used in modeling some parameters; we do not aim at achieving a close quantitative agreement with the literature information [254]. Instead, we seek to obtain values of aggregate quantities that are comparable to those reported by others. We consider three key benchmarking quantities for the validation step, as summarized in Table 4. Based on the acceptable small deviation in this table, our simulation is considered valid.

Table 5 summarizes the verification step for our CFD simulation of the Sakhalin PMHDG channel using the proposed OpenFOAM solver. In this step, we generated an internal benchmarking solution using the same proposed OpenFOAM solver but at a finer spatial resolution and a finer temporal resolution. This auxiliary fine-resolution solution acts as a higher-fidelity reference to assess the nominal (the solution obtained with the nominal resolutions). Based on the observed small changes in the compared characteristic of the MHD channel, as shown in the table, our verification step is considered successful, and our nominal OpenFOAM CFD solution is considered satisfactorily resolution-independent. All the remaining results in this study correspond to the nominal resolutions (which are 376,380 cells and $5 \times 10^{-6}$ s time-step).

**Table 4** Validation of our simulation for the Sakhalin MHD channel

| Quantity | Our value | Benchmarking value | Remarks |
|---|---|---|---|
| Electric power output | 488.1 MW | 476 MW [255] | • While the reported peak power value is 510 MW, we compare our power value with a smaller value (93% of the peak) as an estimated mean value during the pulse. This factor is assigned based on data for the Pamir-3U PMHDG [256] |
| Electric current from electrodes | 191.4 kA | 200 kA [257] | • Our value is the average of the cathode and the anode (area integrated normal component of the electric-current density vectors, perpendicular to the electrode) |
| Plasma speed at channel exit | 1156 m/s | 1306 m/s [258] | • Our value is the mass-averaged axial velocity component (the $X$-component of the plasma velocity vector) at the MHD channel outlet |

**Table 5** Verification of our simulation for the Sakhalin MHD channel (resolution sensitivity analysis)

| Quantity | Nominal resolution | Finer resolution |
|---|---|---|
| Number of cells | 376,380 | 1,235,250 |
| Time step | $5 \times 10^{-6}$ s | $2.5 \times 10^{-6}$ s |
| Electric power output | 488.1 MW | 480.3 MW |
| Volume-average pressure | 3.128 bar | 3.011 bar |
| Volume-average Lorentz force density vector | [−89.12, 28.83, 0] kN/m$^3$ | [−85.86, 28.75, 0] kN/m$^3$ |
| Volume-average electric-current density vector | [1.462, −4.517, 0] A/cm$^2$ | [1.457, −4.351, 0] A/cm$^2$ |

The above verification table also has indications about the validity of our OpenFOAM model, through the correct sign of the $Y$-component of the volumetric Lorentz force density (force per unit volume), which is pointing from the lower cathode to the upper anode for our case where the plasma's bulk motion is in the positive $X$ direction, and the magnetic-field flux density is unidirectional in the positive $Z$ direction. The Lorentz force vector $\vec{F}$ on a general electrically charged particle with a positive charge $q$, when this particle is carried with the plasma gas (moving at the same bulk velocity $\vec{u}$) is [259]

$$\vec{F} = q \left( \vec{E}_L + \vec{u} \times \vec{B} \right) \tag{34}$$

Although the electrostatic field $\vec{E}_L$ reduces or alters the electrodynamic fields $\vec{u} \times \vec{B}$, the resultant overall electric field $\vec{E}_L + \vec{u} \times \vec{B}$ is like electrodynamic fields $\vec{u} \times \vec{B}$ in terms of pointing vertically down (in the negative $Y$ direction, from the upper anode to the lower cathode).

Because the effective mobile charge carriers in the plasma are the negatively-charged electrons (not the positively-charged seed ions), the charge $q$ is actually negative, and thus the Lorentz force actually points upward, and this is in agreement with our CFD predictions [260, 261].

The finding that the averaged electric-current density vector is pointing opposite to the averaged Lorentz force vector is also a favorable indication that the CFD model is able to predict correct vector fields. While the collected conventional electric current flows externally through the electric load from the positive cathode to the grounded anode, the electron electric current flows externally in the opposite direction from the grounded anode to the positive cathode. In the presented OpenFOAM model, the electric-current density refers to the conventional electric current. Therefore, its averaged direction within the MHD channel should be pointing down (thus, having a negative $Y$-component), which is correctly predicted in our simulation.

Finally, the obtained zero $Z$-component of either the averaged Lorentz force density vector or the averaged electric-current density vector is a logical outcome, given the unidirectional $Z$-oriented nature of the applied magnetic-field flux density, and this does not cause any electric-current density or Lorentz force along the $Z$ direction. There is no reason for having an averaged $Z$-component of the electrostatic field (thus, $E_{L,z} = 0$).

## 4 Results

We present the results of the proposed OpenFOAM solver when applied to the Sakhalin magnetohydrodynamic (MHD) channel in the next three subsections.

In the first subsection, we present scalar aggregate quantitative values. In the second subsection, we present one-dimensional graphical profiles. In the third subsection, we present three-dimensional visualizations of selected fields in the entire MHD channel.

### 4.1 Scalar values at the MHD channel outlet

By post-processing our simulation results, we obtain the mass-averaged values of six key selected properties of the plasma as it exits the MHD channel. These values are summarized in Table 6.

A generic mass-averaged quantity (say $\psi$) at the outlet is computed through density-weighted integration of that quantity $\psi_{\text{mass}-\text{avg}}$ at the outlet area, and then dividing the mass flow $\dot{m}$ leaving the channel from that outlet area. Thus,

$$\psi_{mass-avg} = \frac{1}{\dot{m}} \int_{outlet\,area} \rho u \psi dA \tag{35}$$

where $u$ is the axial component of the exiting plasma (this component is perpendicular to the outlet surface), $dA$ is an infinitesimal area element at the outlet (which has a total finite area of 1.6 m$^2$).

To better demonstrate the change in the plasma flow due to the direct power extraction (DPE) process within the MHD channel, the inlet values in our OpenFOAM simulation for the same six plasma properties are also listed in the table, which helps in contrasting them with the computed outlet values.

It can be seen that one major influence of the power extraction process from the plasma is an occurring deceleration process, although supersonic flows accelerate in divergent channels (as in the case of the Sakhalin channel) [262, 263].

**Table 6** Computed properties of the plasma (mass-averaged) at the channel outlet, and corresponding inlet values

| Quantity | Outlet value | Inlet value |
| --- | --- | --- |
| Absolute temperature (K) | 2,738.4 | 2,750 |
| Absolute pressure (bar) | 3.294 | 3.28 |
| Axial velocity (m/s) | 1,156 | 2,050 |
| Mach number (-) | 1.161 | 2.051 |
| Electric conductivity (S/m) | 48.4 | 50 |
| Hall parameter (-) | 0.2555 | 0.2715 |

On the other hand, the temperature and the pressure nearly did not change from their inlet values. This means that the main source of the electricity extracted from the flowing plasma is its kinetic energy, rather than its internal energy (thermal energy) or its pressure energy (flow energy).

Due to the weak change in the pressure and temperature, and because the electric conductivity and the Hall parameter are fitted as dependent functions of these two thermodynamic properties (the Hall parameter depends additionally on the magnitude of the applied magnetic-field flux density), the electric conductivity and the Hall parameter of the plasma do not change largely when the outlet location is compared with the inlet location.

Because the Mach number depends largely on the axial velocity of the plasma, and because the speed of sound is dependent on the temperature which does not show a large change between the inlet and the outlet locations of the MHD channel; the Mach number decreases strongly between the inlet and the outlet locations, similar to the change in the axial plasma velocity. This decrease of a supersonic Mach number for a flow within a divergent channel or nozzle also emphasizes the loss of kinetic energy from the plasma, which is actually decelerated within the channel because of the electric power extraction.

### 4.2 Centerline profiles

Through one-dimensional sampling along the centerline of the MHD channel (thus, along all $X$ values from 0 to 4.5 m, but with $Y = 0$ and $Z = 0$), the variations of ten thermodynamic, fluidic, or electric properties of the simulated Sakhalin channel are illustrated in the current subsection. None of the investigated profiles has a discontinuity (abrupt change), which is an implicit indication of the lack of internal instabilities in the performed simulation, which agrees with the expectation for this problem and supports the correctness of the presented model.

We start with the electric conductivity $\sigma$, and display its centerline variation in Fig. 5. We point out here that this plasma electric variable is described through an empirical nonlinear regression model as a function of the local temperature and the local pressure. Therefore, its variation reflects variation in either of these two independent variables. Despite the apparent non-uniformity of the electric conductivity, its variation is smooth and bounded within a narrow range of about 4 S/m only (from 48.6 S/m to 52.8 S/m).

Next, we display the variations of the Hall parameter $\beta$ along the centerline of the MHD channel in Fig. 6. Like the electric conductivity, the Hall parameter is described through an empirical nonlinear regression model as a function of the local temperature and the local pressure. In addition, the Hall parameter depends linearly on the magnitude of the applied magnetic-field flux density $B$. The magnitude of the applied magnetic-field flux density declines rapidly near the inlet and outlet of the MHD channel, and this is clearly shown in the profile of the Hall parameter. In the internal region of the channel, the Hall parameter qualitatively follows a similar profile to that of the unidirectional applied magnetic-field flux density, which can be explained by the weak influence of the temperature and the pressure because they do not change significantly within the MHD channel. The maximum encountered value of the Hall parameter is 0.39, while the lowest value is at the MHD outlet, and it is 0.26.

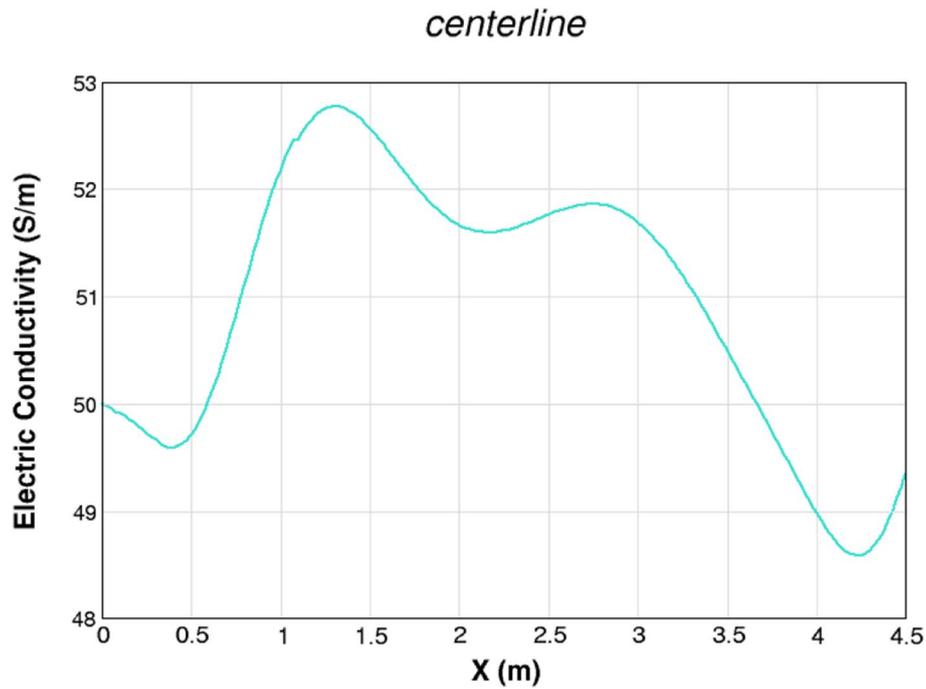

**Fig. 5** Centerline profile of the electric conductivity along the simulated Sakhalin MHD channel

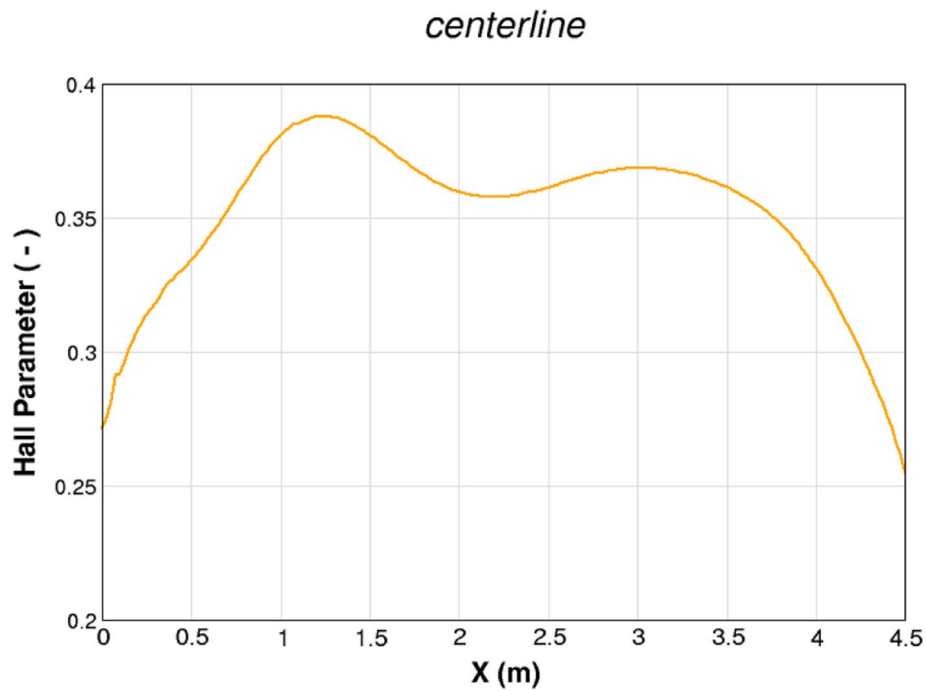

**Fig. 6** Centerline profile of the Hall parameter along the simulated Sakhalin MHD channel

Figure 7 shows the centerline profile of the Mach number, which declines in a nearly monotonic pattern, leaving the MHD channel at a noticeably smaller value of 1.26 than its inlet value of 2.05. However, the plasma remains supersonic even at the outlet (the exit Mach number is still above unity). Therefore, the existing plasma still has a lot of kinetic energy that can be exploited in a subsequent energy conversion system.

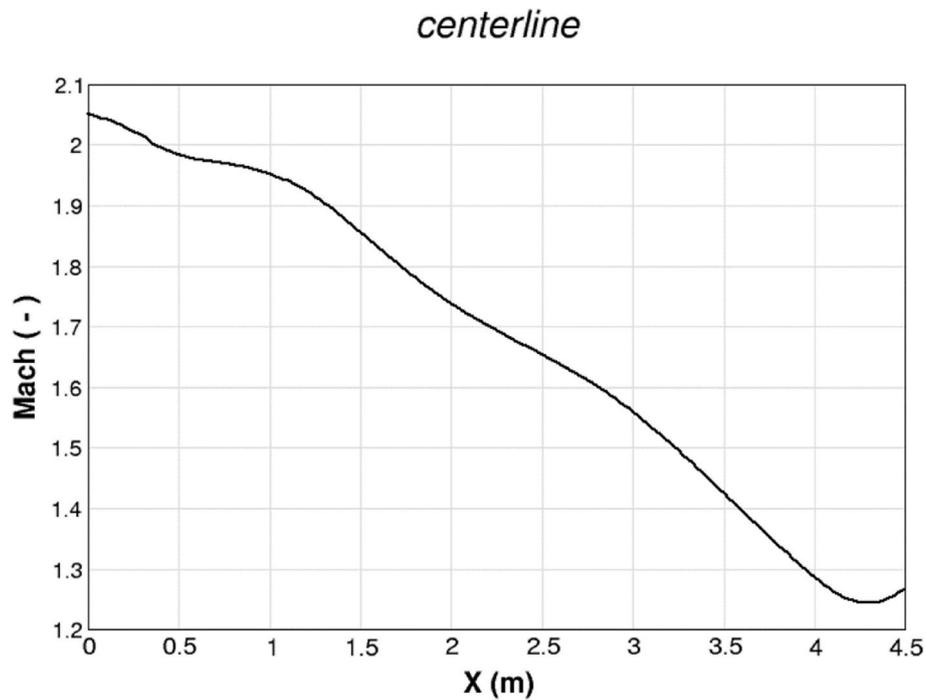

**Fig. 7** Centerline profile of the Mach number along the simulated Sakhalin MHD channel

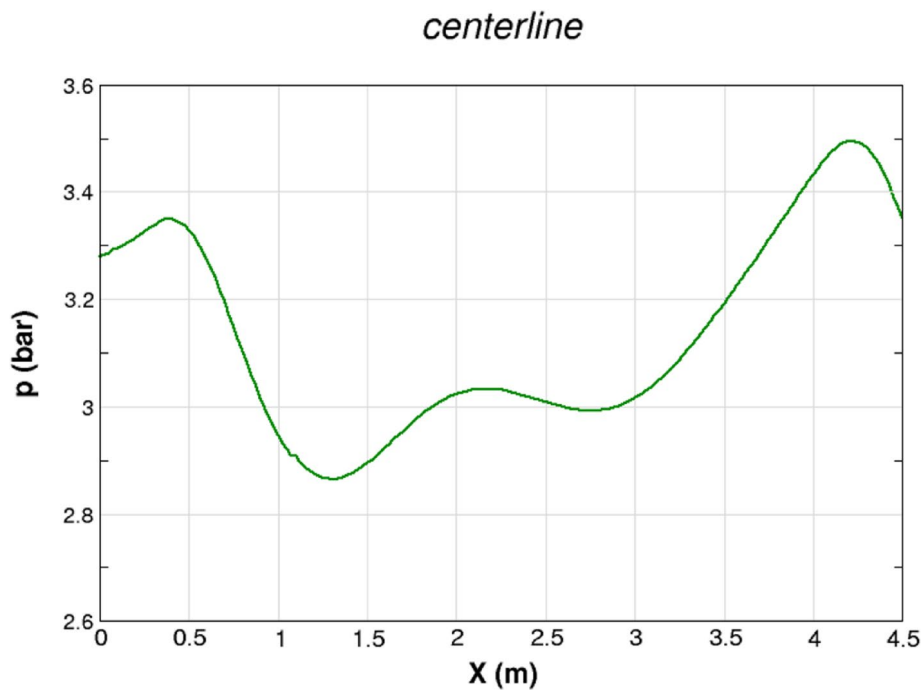

**Fig. 8** Centerline profile of the absolute pressure along the simulated Sakhalin MHD channel

Figure 8 shows the centerline profile of the absolute pressure, which manifests slow variations within 0.6 bar. The leaving plasma still has a high pressure close to the inlet pressure (in addition to the still-high speed), thus it is subject to further energy extraction through another energy system (such as gas turbines) after deceleration to proper subsonic speeds using supersonic diffusers [235, 264].

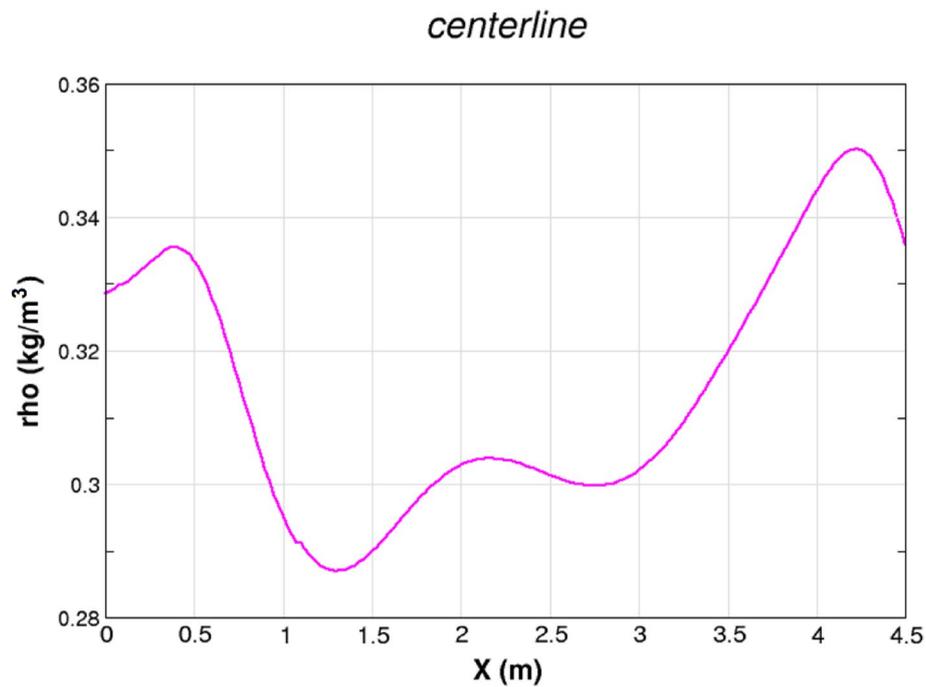

**Fig. 9** Centerline profile of the plasma density along the simulated Sakhalin MHD channel

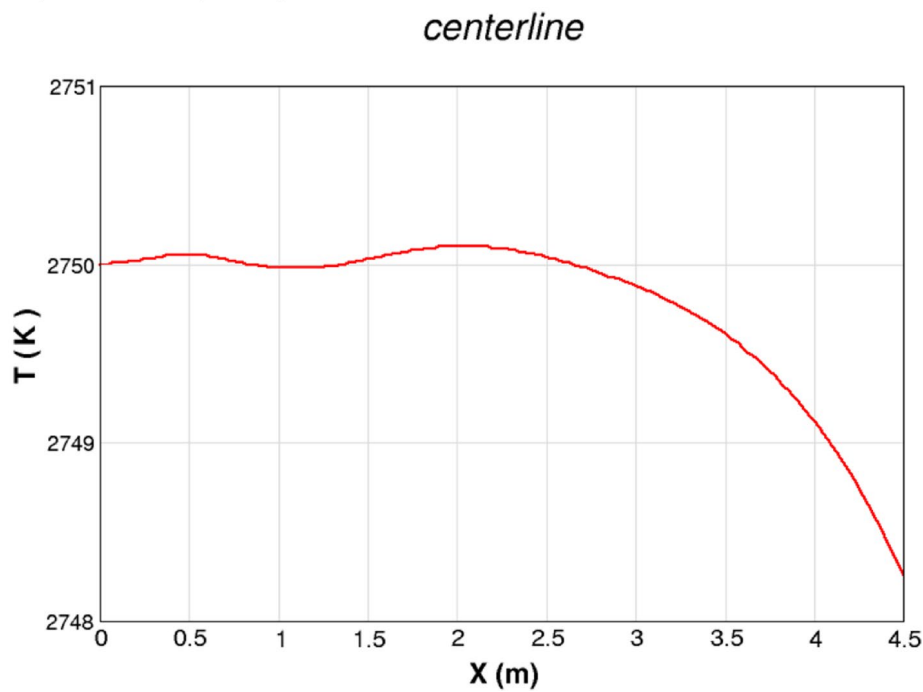

**Fig. 10** Centerline profile of the absolute temperature along the simulated Sakhalin MHD channel

Figures 9 and 10 show the centerline profiles of the plasma density and the plasma absolute temperature, respectively. It is interesting that the temperature is practically constant near its inlet value of 2,750 K. Following the ideal gas law, the density in such a case (a nearly isothermal process) becomes essentially a scaled version of the absolute pressure, which is a satisfied feature that can be confirmed by comparing the previous profile of the absolute pressure with the profile of the density. The high temperature of the leaving plasma indicates an enormous amount of thermal energy that can be transferred through a heat exchanger to another fluid, for example, to generate steam that can be used in a steam power cycle. Using the empirical formula for the specific heat capacity at constant pressure for the plasma of the Sakhalin PMHDG, namely.

$C_p(T[K])[J/(kg.K)] = 649.863 + 0.620932T$, the change in the specific enthalpy, designated by.

$\Delta h \equiv h_2 - h_1$, between two states (a lower-temperature state 1 and a higher-temperature state 2) can be obtained by integrating this $C_p$ expression as

$$\Delta h \equiv h_2 - h_1 = \int_{T_1}^{T_2} C_p(T)\,dT \tag{36}$$

This gives

$$\Delta h = 649.863(T_2 - T_1) + 0.310466(T_2^2 - T_1^2) \tag{37}$$

If the higher-temperature state has $T_2$ = 2,750 K (nearly the plasma exit temperature), and the lower-temperature state has $T_1$ = 300 K (nearly the ambient temperature), then the enthalpic energy available per kg of plasma is $3.912 \times 10^6$ J/kg (3.912 MJ/kg). This is more than enough to generate superheated steam at extremely high temperatures (even exceeding 700 °C), which is suitable to operate steam power plants [265, 266].

Figures 11 and 12 show the centerline profiles of the axial and upward velocity components of the plasma, respectively. As was the case for the Mach number, the centerline axial velocity declines as the plasma progresses through the MHD channel, reflecting a loss of kinetic energy that is converted into electric energy. The upward velocity component (the plasma velocity in the $Y$ direction, being positive if the plasma is moving from the downward cathode to the upward anode) varies mildly around the zero value, bounded between −50 m/s and 30 m/s.

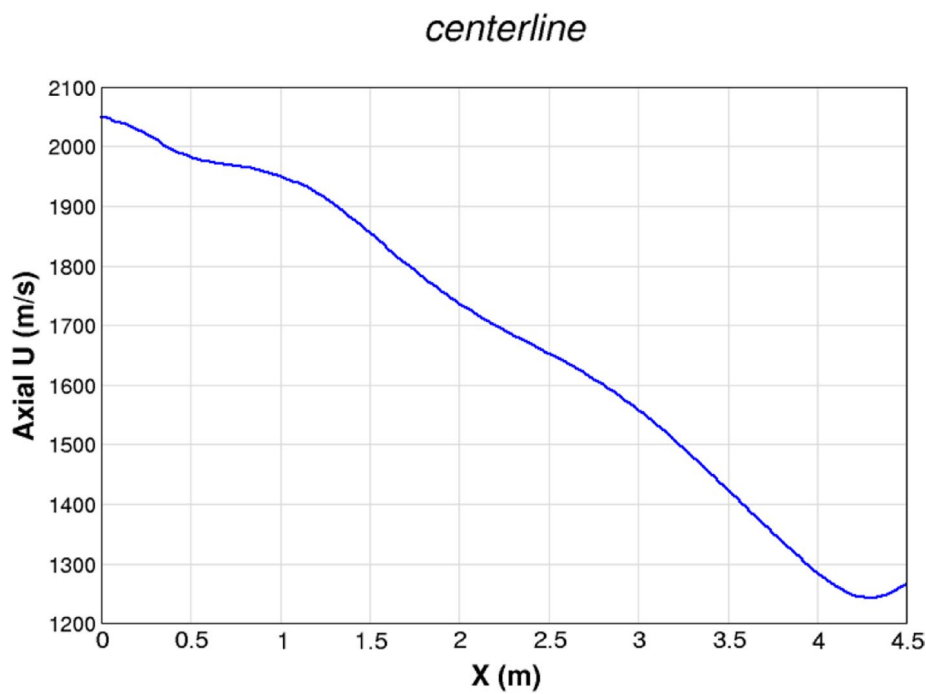

**Fig. 11** Centerline profile of the axial velocity component along the simulated Sakhalin MHD channel

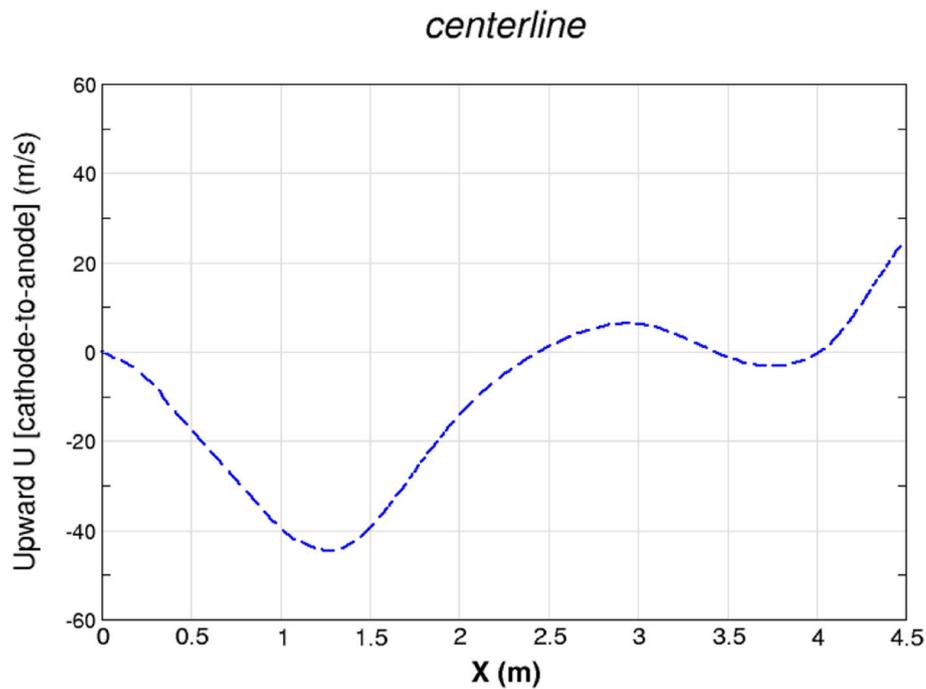

**Fig. 12** Centerline profile of the upward velocity component (the $Y$-compoent) along the simulated Sakhalin MHD channel

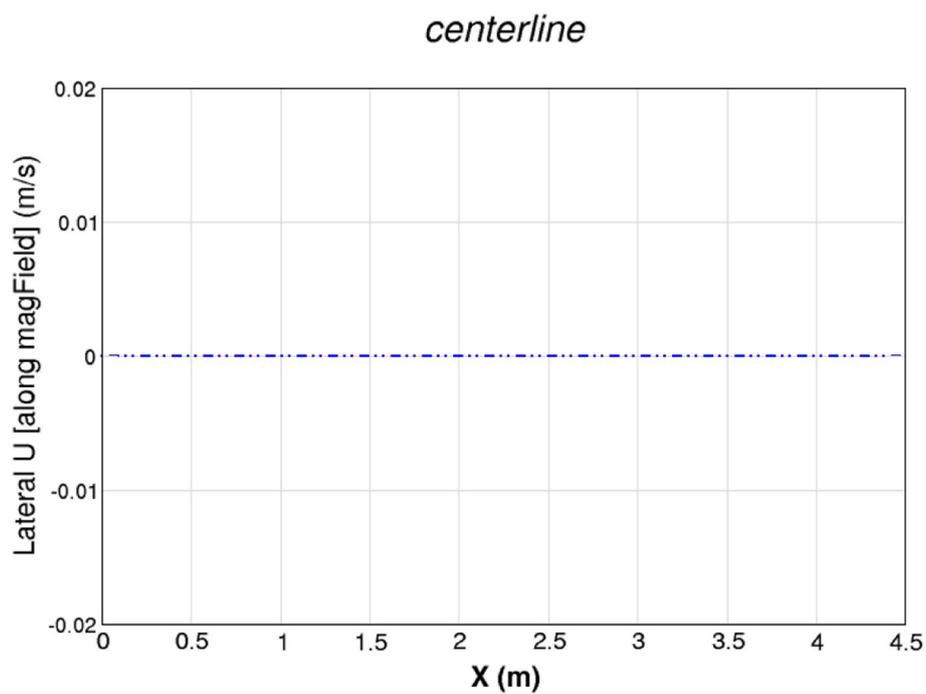

**Fig. 13** Centerline profile of the lateral velocity component (the $Z$-compoent) along the simulated Sakhalin MHD channel

Figure 13 shows the centerline profile of the third velocity component of the plasma (the $Z$-component), which is along the unidirectional applied magnetic-field flux density. This component is practically zero at the centerline, reflecting perfect symmetry and balance of lateral forces. This is reasonable given the lack of gravitational effect along the $Z$ direction, as well as the constant channel width along the same direction.

Figure 14 shows the centerline profile of the downward component of the electric-current density. This component has negative values; thus, the conventional electric-current density is directed toward the bottom cathode (although the physical charge-carrying electrons are forced upward toward the upper anode). In the figure, the negative value of that already-negative conventional component is displayed, which is equivalent to displaying the absolute value of that negative conventional component. The displayed absolute conventional current density component increases nonlinearly until reaching a maximum value of 6.1 A/cm$^2$ at a distance of 1 m from the MHD channel inlet; it then declines nonlinearly until reaching a zero value at the channel outlet.

### 4.3 Three-dimensional distributions

In this subsection, we present three-dimensional views for eight selected fields as per the solution obtained in our simulation of the Sakhalin MHD channel. These views are instrumental in demonstrating the variation of each field in the entire channel.

Figure 15 displays the spatial distribution of the electric conductivity, which changes mildly between 43.9 S/m and 53.6 S/m. The electric conductivity declines slowly as the plasma approaches the rear end of the divergent channel. It is worth mentioning here that this electric conductivity (being near 50 S/m) is roughly ten times the electric conductivity of seawater (being approximately 5 S/m) [267]. Additionally, the estimated range of the maximum electric conductivity of Earth's lower mantle is 3–10 S/m [268].

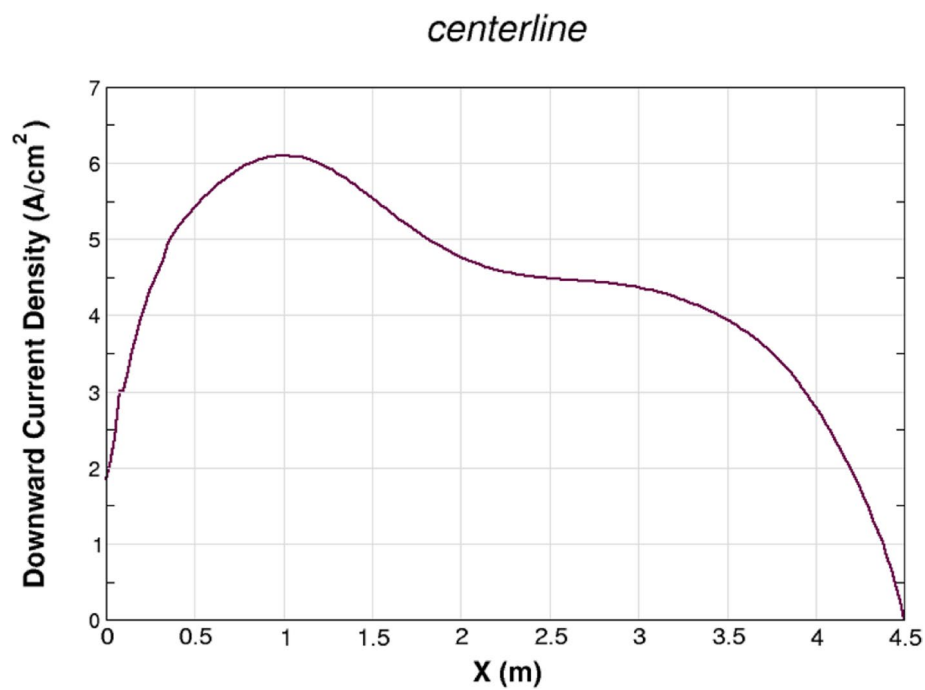

**Fig. 14** Centerline profile of the downward component of the electric-current density (from anode to cathode) along the simulated Sakhalin MHD channel

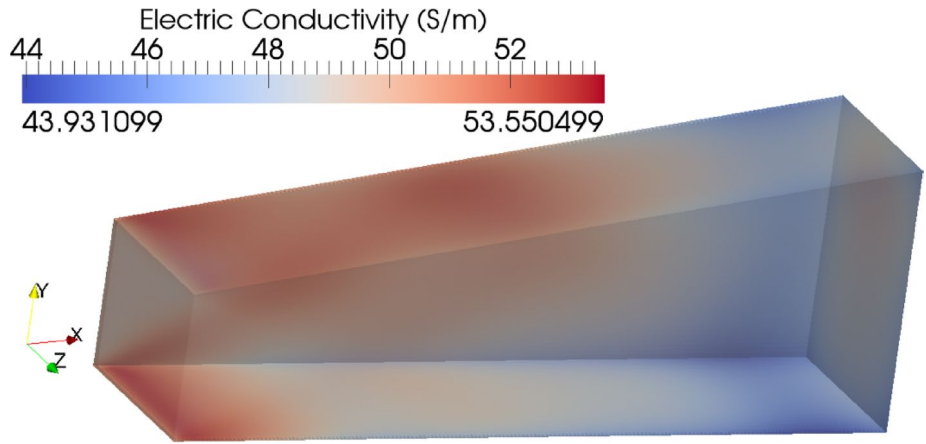

**Fig. 15** Three-dimensional distribution of the electric conductivity within the simulated Sakhalin MHD channel

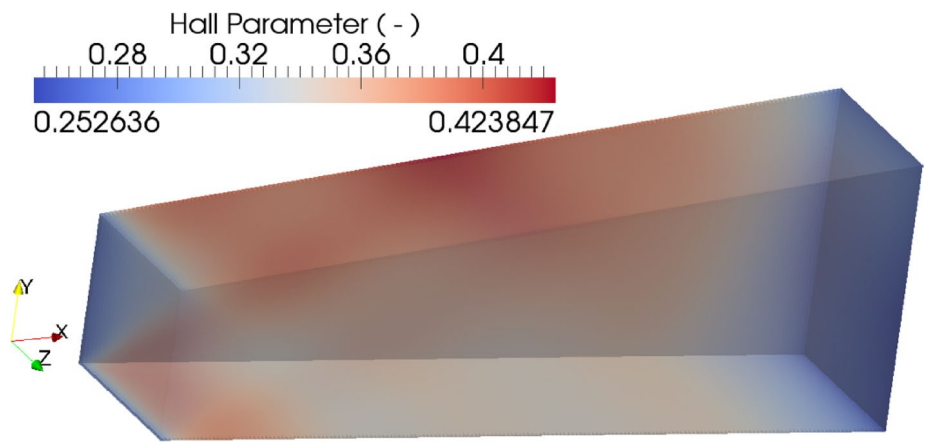

**Fig. 16** Three-dimensional distribution of the Hall parameter conductivity within the simulated Sakhalin MHD channel

Figure 16 displays the spatial distribution of the Hall parameter, which changes between 0.253 and 0.424. The variations in the Hall parameter are determined by variations in the temperature, pressure, and the applied magnetic-field flux density.

Figure 17 displays the spatial distribution of the absolute pressure, which changes between 2.56 bar and 3.62 bar. A pressure recovery phenomenon is observed near the outlet, where the pressure increases back after its initial drop after entering the channel.

Figure 18 displays the spatial distribution of the axial velocity component of the plasma, which changes between 832.9 m/s and 2,089 m/s. The deceleration process is efficiently visible through this figure, with the axial velocity generally dropping as the plasma moves from the inlet toward the outlet.

Figure 19 displays the spatial distribution of the specific turbulent kinetic energy (the turbulent kinetic energy per unit mass, designated by the symbol $k$), which changes between 1,803 $m^2/s^2$ and 16,262 $m^2/s^2$. This variable represents the intensity of the modeled turbulence in the plasma. This turbulence quantity is higher in the intermediate part of the channel compared to the inlet and the outlet.

Figure 20 displays the spatial distribution of the turbulence dissipation or the turbulence dissipation rate (designated by the symbol $\epsilon$ or the name "epsilon"), which changes between $2.05 \times 10^5$ $m^2/s^3$ and $7.2 \times 10^7$ $m^2/s^3$. This variable describes the loss of turbulent motion that is converted into thermal internal heat due to viscous effects [269]. This turbulence quantity is higher in the intermediate part of the channel than in the inlet and outlet. This was the case for the turbulent kinetic energy per unit mass, and it is reasonable behavior that as the turbulence intensity increases, the turbulence dissipation also increases [270].

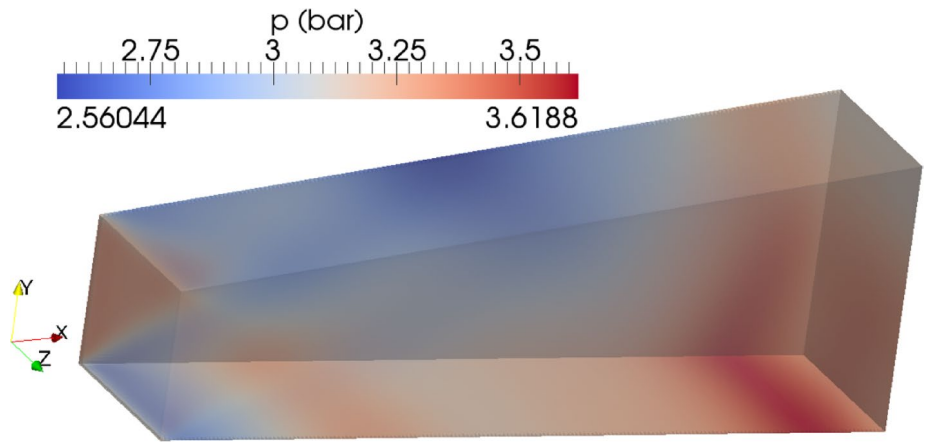

**Fig. 17** Three-dimensional distribution of the absolute pressure within the simulated Sakhalin MHD channel

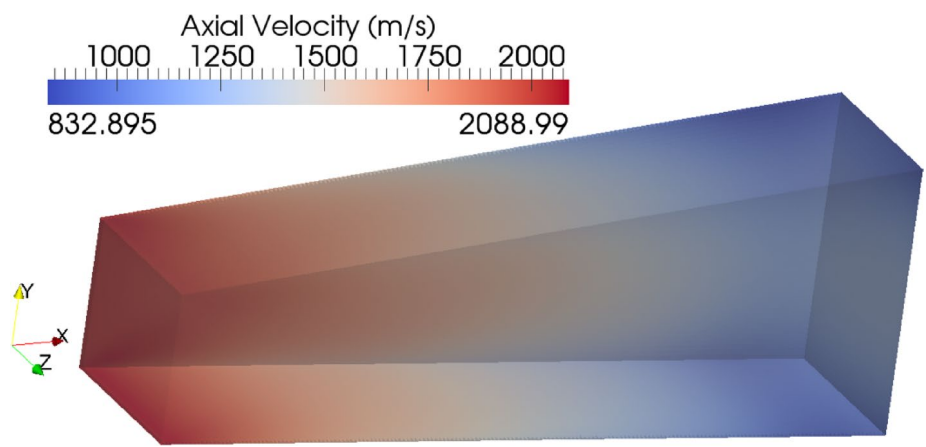

**Fig. 18** Three-dimensional distribution of the axial velocity component within the simulated Sakhalin MHD channel

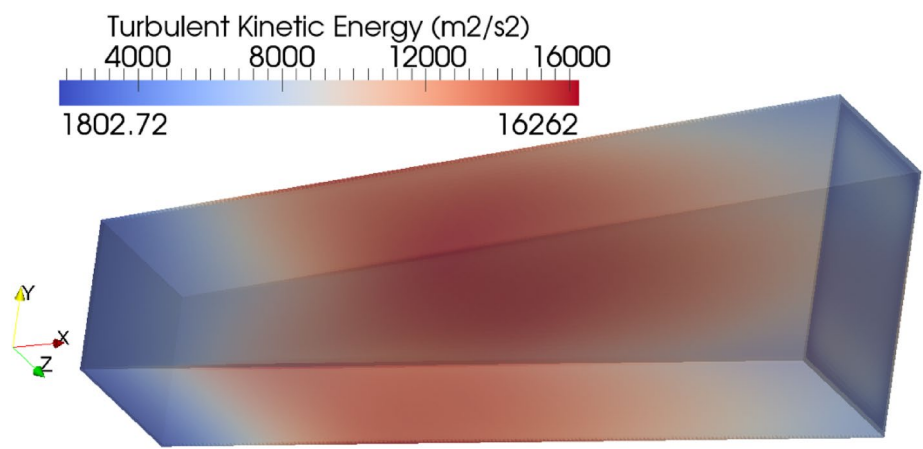

**Fig. 19** Three-dimensional distribution of the specific (per unit mass) turbulent kinetic energy within the simulated Sakhalin MHD channel

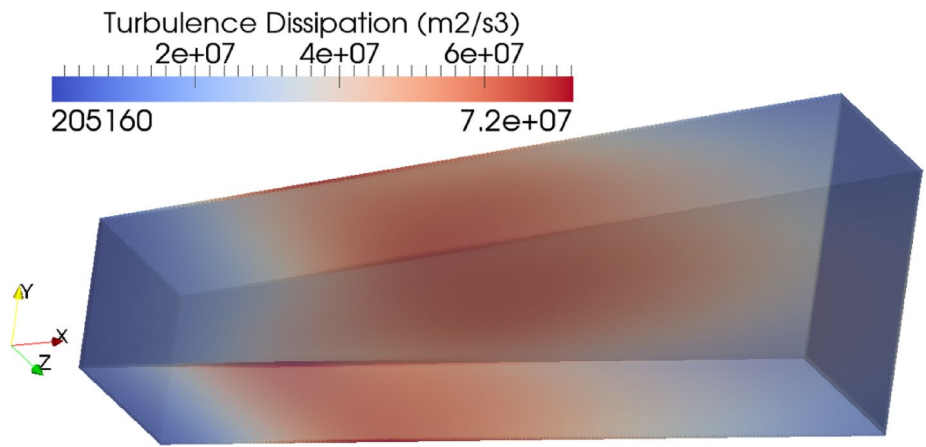

**Fig. 20** Three-dimensional distribution of the dissipation (epsilon) of the turbulent kinetic energy within the simulated Sakhalin MHD channel

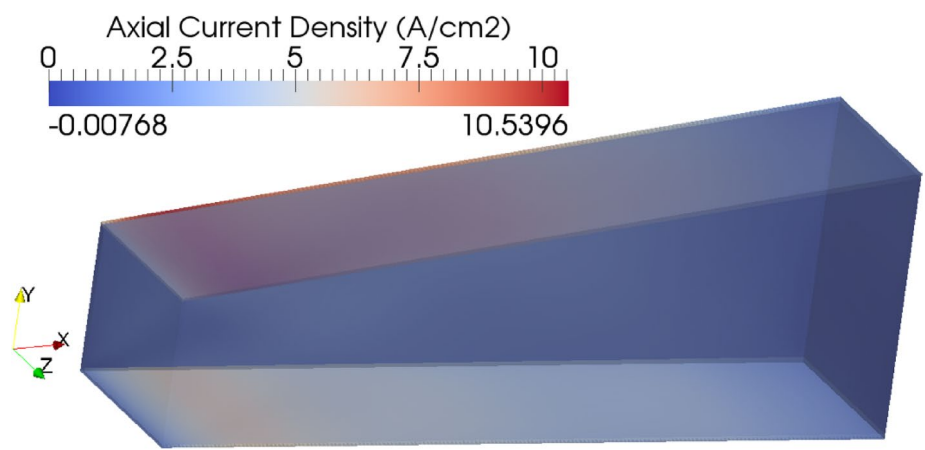

**Fig. 21** Three-dimensional distribution of the axial component of the electric-current within the simulated Sakhalin MHD channel

Figures 21 and 22 display the spatial distribution of the axial and downward components of the electric-current density, respectively. Thus, the variables displayed in these figures are $J_x$ and $-J_y$, respectively. The axial component changes between −0.008 A/$cm^2$ and 10.5 A/$cm^2$. Thus, the axial electric-current density is mostly from the channel inlet to the channel outlet. The downward component changes between −0.310 A/$cm^2$ and 21.9 A/$cm^2$. Thus, the vertical electric-current density is mostly from the top anode to the bottom cathode.

## 5 Conclusions
### 5.1 Concluding remarks

In the current study, we presented a mathematical and computational model for simulating linear magnetohydrodynamic (MHD) Faraday channels with continuous electrodes.

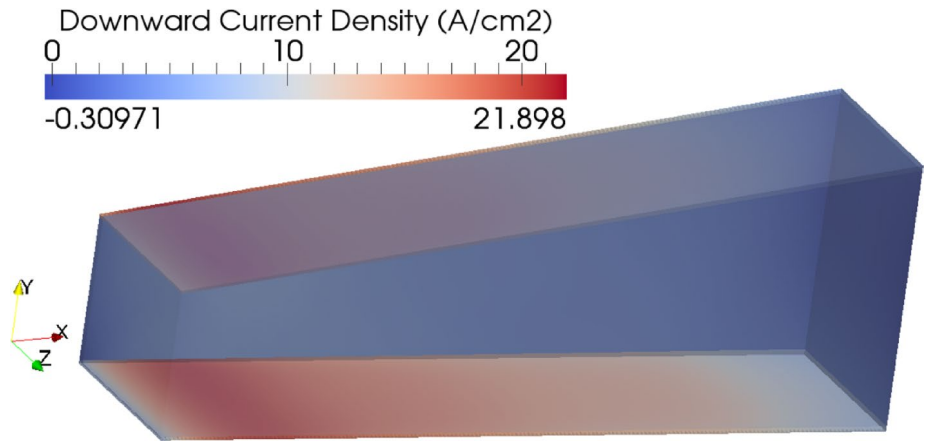

**Fig. 22** Three-dimensional distribution of the downward component of the electric-current density within the simulated Sakhalin MHD channel

The computational fluid dynamics (CFD) computer language OpenFOAM is used to develop an MHD solver, which we applied to the Sakhalin pulsed solid-propellant plasma (SPP) power system, with a supersonic divergent channel. The proposed solver corresponds to the low magnetic Reynolds number ($Re_m$), and it utilizes the scalar electric potential as the working variable for handling the electric aspect of the problem.

After validation and verification, we used the solver to explore various aspects of the thermal, fluidic, and electric features of the problem. These include the pressure, temperature, density, velocity, Mach number, electric-current density, specific turbulent kinetic energy, and turbulence dissipation.

## 6 Summary of contributions

Despite earlier studies about this Sakhalin pulsed magnetohydrodynamic generator (PMHDG), we do not know of any previous work in the literature that provided the same comprehensive aspects as our study.

The following items can be considered as contributions made in this study:

- A review of real examples of pulsed MHG generators (PMHD) was made, with emphasis on the Sakhalin model that was described with quantitative values and empirical regression models for its operational settings that help in modeling it by others. A large number of literature research articles were used in preparing this review.
- A mathematical/computational model is proposed for simulating the three-dimensional fluid dynamic and electric fields for linear continuous-electrode MHD channels. The proposed solver was found to yield satisfactory outputs when compared to other independent studies. The ability of the solver to give resolution-independent results was verified.
- Selected results from our CFD simulation for the Sakhalin MHD channel were presented and discussed. Using the power of CFD and data post-processing, it was possible to reveal various attributes of the channel of the Sakhalin PMHDG, such as the nearly-isothermal operation, the extraction of energy mainly from the kinetic energy of the plasma (rather than the thermal or the pressure energies), the unusual deceleration (opposite to the conventional situation) of supersonic flow despite flowing in a divergent channel, the weak change in the electric conductivity, and the pressure recovery phenomenon.

### 6.1 Possible future research

The MHD solver presented here can be expanded in subsequent studies to make it even more useful as a generic tool for designing MHD channels for open-cycle MHD generators, without being customized to the Sakhalin channel.

One way to improve the applicability of the presented model is to replace the prescribed regression function for the local plasma electric conductivity with a submodel that estimates the local plasma electric conductivity for an arbitrary thermal-equilibrium plasma medium, given its (1) chemical composition, (2) temperature, and (3) pressure.

Another way to extend this study is to perform sensitivity analyses of channel geometry parameters (such as the divergence angle and electrode length) to identify their influence and the opportunity to optimize the MHD channel.

One more way to improve the OpenFOAM MHD solver presented here is to address one or more of its limitations mentioned earlier.

**Abbreviations**
| | |
|---|---|
| CAD | Computer-aided design |
| CCGT | Combined-cycle gas turbine |
| CFD | Computational fluid dynamics |
| DC | Direct current |
| DPE | Direct power extraction |
| ECE | External combustion engine |
| FANS | Favre-averaged Navier–Stokes equations |
| FSI | Fluid–structure interaction |
| FVM | Finite volume method |
| GHG | Greenhouse gas |
| HRSG | Heat recovery steam generator |
| ICE | Internal combustion engine |
| MHD | Magnetohydrodynamics |
| OFCC | Oxy-fuel carbon capture |
| OOP | Object-oriented programming |
| PG | Plasma generator |
| PMHDG | Pulsed magnetohydrodynamic generator |
| PV | Photovoltaic |
| RANS | Reynolds-averaged Navier–Stokes equations |
| SPP | Solid-propellant plasma |
| TEG | Thermoelectric generator |
| USSR | Union of Soviet Socialist Republics |

**Author contributions**
O.M. is the single author of this manuscript.

**Funding**
Not applicable (this research received no funding).

**Data availability**
Data supporting this study is provided within the manuscript.

**Declarations**

**Ethics approval and consent to participate**
Not applicable.

**Consent for publication**
Not applicable.

**Competing interests**
The author declares that they have no known competing financial interests or personal relationships that could have appeared to influence the work reported in this paper.

Received: 26 February 2025 / Accepted: 6 September 2025

Published online: 30 September 2025

**Publisher's Note**

Springer Nature remains neutral with regard to jurisdictional claims in published maps and institutional affiliations.